\documentclass[12pt]{iopart}
\usepackage{epsfig}  
\usepackage{pstricks}
\usepackage{rotate}

\begin{document}

\newcommand{\gapprox}{\stackrel{>}{_{\sim}}}
\newcommand{\lapprox}{\stackrel{<}{_{\sim}}}
\newcommand{\ra}{\rightarrow}
\newcommand{\ccb}{c\bar{c}}
\newcommand{\bbb}{b\bar{b}}
\newcommand{\pbp}{\bar{p}p}
\newcommand{\epem}{e^+e^-}
\newcommand{\gaga}{\gamma\gamma}
\newcommand{\ftc}{F_2^{c}}
\newcommand{\ptr}{p_T^{rel}}
\newcommand{\pbmo}{pb$^{-1}$}

\title[Open Beauty Production]
      {Open Beauty Production}

\author{Felix Sefkow
\footnote{On behalf of the H1 and ZEUS Collaborations.}
}

\address{Physik-Institut der Universit\"at Z\"urich \\
Winterthurerstr.\ 190, CH-8057 Z\"urich, Switzerland}

\begin{abstract}
We review measurements of open beauty production at HERA,
with emphasis on recent results based on lifetime signatures. 
The beauty cross sections in photoproduction and deep-inelastic
scattering are found to be higher than expected 
in QCD at next-to-leading order. 
The discussion includes new results on beauty production
in $\epem$, $\gaga$ and $\pbp$ interactions. 
An outlook on the potential for measurements 
with the upgraded HERA collider and experiments is also given. 
\end{abstract}




\section{Introduction}

Almost 25 years after the discovery of the $b$ quark 
in proton nucleus collisions~\cite{bdisc},
the accurate understanding of how 
$b$ quarks are being produced in hadronic environments 
is still an open issue.  
The subject of beauty production has recently received renewed interest, 
with the advent of 
new measurements at the $ep$ collider HERA, and elsewhere. 
 
The dominant mechanism for the production of heavy quarks at HERA 
is photon gluon fusion: a photon coupling to the scattered electron
interacts with a gluon from the proton by forming a quark antiquark 
pair, e.g.\ $\bbb$. 
A quantitative description of the process requires the 
knowledge of the gluon momentum distribution in the proton, the 
calculation of the hard photon gluon subprocess, and 
a fragmentation function  which accounts 
for the long-range effects binding the heavy quark in 
a hadron.
The gluon density in the relevant range of momentum fraction $x$
and at the appropriate scales is known 
to an accuracy of a few percent from the analysis of 
scaling violations of the proton structure function $F_2$ 
measured at HERA~\cite{f2h1zeus}. 
The partonic process has been calculated in QCD, at next-to-leading order
(NLO).  The masses of the charm, and even more so of the beauty quark 
ensure that at least one hard scale is present 
that renders QCD perturbation theory to be applicable.
The fragmentation function is extracted from $\epem$ annihilation data, 
where the kinematics of the hard process is well determined 
by a clean initial state.
New results on $b$ fragmentation, with significantly improved precision,
have appeared recently~\cite{alephbfrag,sldbfrag} and are
included in this review. 
With such precise ingredients, $b$ production at HERA provides a QCD testing 
ground {\it par excellence}.
The validity range of the perturbative methods and the universality
of non-perturbative phenomenological inputs can be closely examined. 

Understanding heavy quark production is furthermore essential
for the proton structure analysis of inclusive deep-inelastic scattering (DIS) 
data~\cite{f2h1zeus} in terms of parton distribution functions, 
since final states with charm account for about a quarter of the 
inclusive rate in the kinematic domain probed at HERA~\cite{f2ch1zeus}. 
As shown in more detail in~\cite{redondo},
the QCD picture sketched above has proven very successful in describing  
the experimental results on charm production in DIS~\cite{f2ch1zeus,h1glue}.
The limit of small photon virtuality is equivalent to charm 
photoproduction.
Even in this regime, 
the picture works reasonably well~\cite{h1glue,h1gpcharm,zeusgpcharm},
albeit some further clarification is needed with regard to 
the treatment of resolved photon processes,
where the photon flutuates into a hadronic state 
which interacts with the proton. 
Evidently, it is of particular interest to subject the theory to an independent
test with beauty production data.

The NLO QCD calculations of heavy quark production 
follow two basic approaches; 
for a more profound discussion see~\cite{kramer}.  
The cross section for e.g.\ photoproduction of heavy hadrons
factorizes and can be expressed as 
convolutions 
of parton distribution functions for photon and proton, 
$f_i^{\gamma,p}(x_{\gamma,p})$, respectively, 
the partonic cross section 
$\hat{\sigma}_{i,j}$ ($i,j$ denoting the parton type),
and a non-perturbative fragmentation function $D(z)$,
which maps the subsequent transition 
of the produced heavy quark into an observable heavy hadron, 
retaining a fraction $z$ of the heavy quark's momentum, symbolically  
\begin{equation}
\sigma_{\gamma p} = 
\sum_{i,j}f_i^{\gamma}(x_{\gamma}) \;\otimes\; f_j^{p}(x_{p}) \;
\otimes\; \hat{\sigma}_{i,j} \;\otimes\; D(z) \; .
\end{equation}
In the so-called massive scheme, 
only light quarks and gluons are active partons 
in the initial state.            
The heavy quark mass $m_Q$ sets the scale for the perturbative expansion
of $\hat{\sigma}$ which has been evaluated up to ${\cal O}(\alpha_s^2)$, 
including mass effects. 
If a second and different large scale is present, 
e.g.\ at large transverse momentum of the heavy quarks, $p_T\gg m_Q$, 
these calculations 
acquire large logarithms of the ratio $p_T/m_Q$,
and the  fixed order result becomes less reliable.  
In ``resummed'' calculations~\cite{kniehlhera,cacciarihera},
in the so-called massless approach, the leading terms of this type
are absorbed in scale-dependent fragmentation functions, 
which obey the Altarelli-Parisi evolution equations.  
Different schemes exist for the construction of fragmentation functions. 
The heavy quarks are treated 
in the same way as the light quarks. 
They appear in the initial state structure functions for proton and photon 
and contribute via flavour excitation processes to the production 
of heavy quark final states.  
However, mass effects in the partonic cross section are neglected.
The massive and the massless approach thus have complementary ranges 
of applicability -- low and high $p_T$ -- which do not necessarily overlap. 
Recently, ``merged'' calculations interpolating between the schemes have 
appeared~\cite{friximerged}.    
Only in the massless scheme, 
the factorization theorem guarantees
the universality of the fragmentation functions extracted from 
$\epem$ data.  
For the scale-independent function $D(z)$ in the massive scheme, 
this remains an assumption based on less rigorous arguments. 
 
The HERA results for beauty are so far limited by small statistics 
and dominated by production near threshold. 
The calculations supposed to be appropriate in this regime 
follow the fixed order `massive' approach 
and are available in the form of Monte Carlo integration programs
for photoproduction~\cite{fmnr} and DIS~\cite{hvqdis}.
Due to the higher quark mass, the QCD predictions are expected to 
be more reliable for beauty than for charm. 
However, we note that the NLO corrections to the predicted
HERA cross section are around 40\% of the LO result in both cases. 

In the next section the measurements performed at HERA
will be presented in detail and confronted with theoretical
expectations. 
In a subsequent section, results from other production environments 
will be summarized. $\epem$ data on fragmentation and 
measurements of beauty production cross sections 
in $\pbp$ and $\gaga$ collisions will 
be presented and compared with QCD calculations 
which follow the same principles as for $ep$ collisions.     
In the last part, we will come back to HERA and discuss the potential 
offered by the HERA upgrade programme~\cite{h1upgrade,zeusupgrade} 
for further studies in the field.

\section{Measurements of beauty production at HERA}

Beauty production at HERA is suppressed 
by two orders of magnitude with respect to charm, 
due to the larger mass and smaller electric charge of the $b$ quark. 
The total cross section is dominated by photoproduction.
Final states are characterized by a steeply falling $p_T$ spectrum, 
which represents a challenge for secondary vertex detection. 
With $m_b/m_c\sim 3$, the minimally required momentum fractions 
of the initial state gluon
is about $10$ times higher than for charm production;
typical values of $x$ are around $10^{-2}$ 
and extend up to a few times $10^{-1}$. 
The corresponding boosts relative to the laboratory frame 
are such that most $b$ quarks produced at HERA are 
emitted under central rapidities into detector regions well 
covered by existing~\cite{cst} or to be commissioned~\cite{zeusupgrade} 
Silicon tracking devices. 

\begin{figure}[h]
  \begin{picture}(28,13)(-18,-3)
\put(-18,-3.2){\epsfig{file=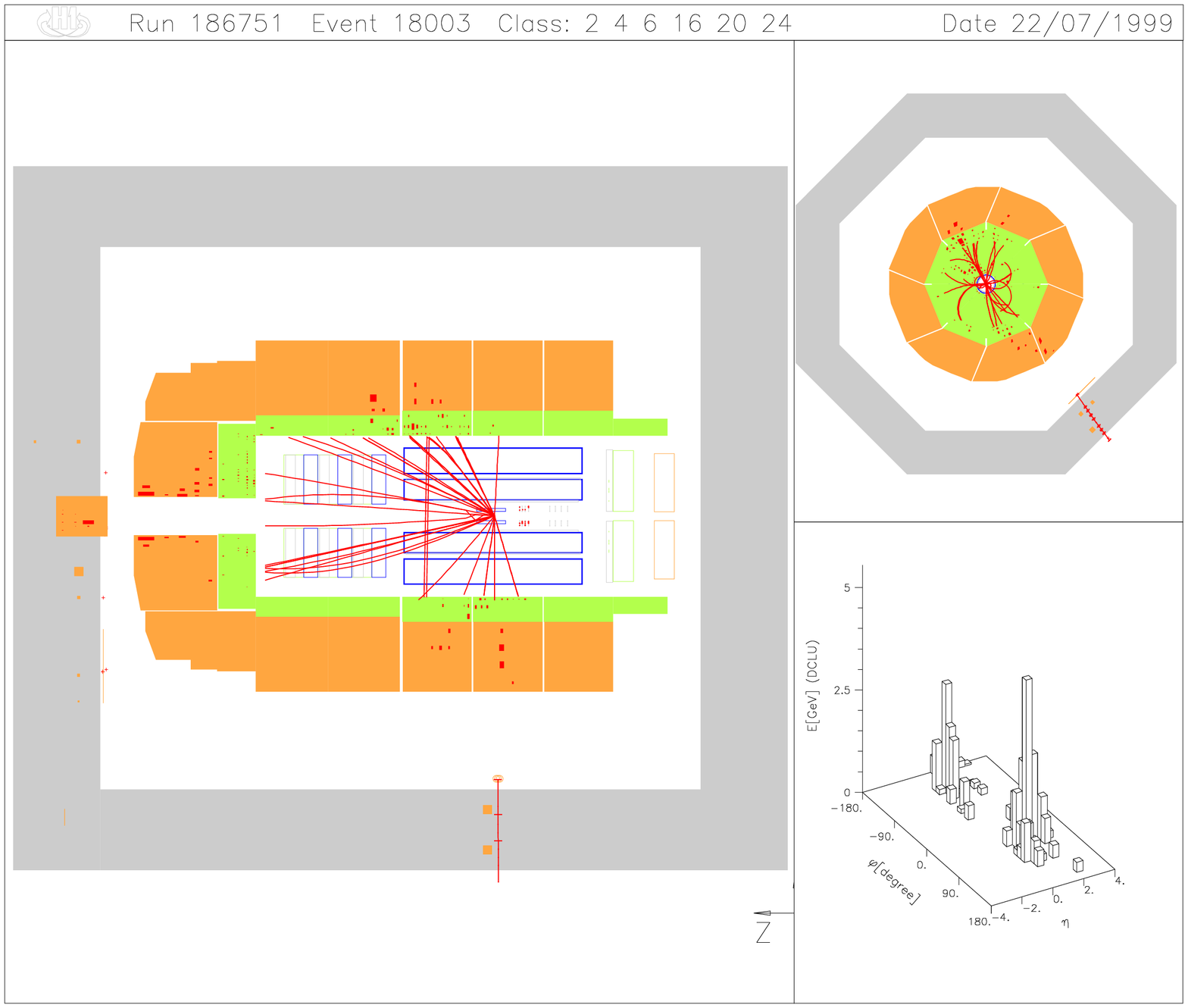,width=7.2cm,angle=90}}
  \put(0.0,0.0){\epsfig{angle=0,width=5.4cm,file=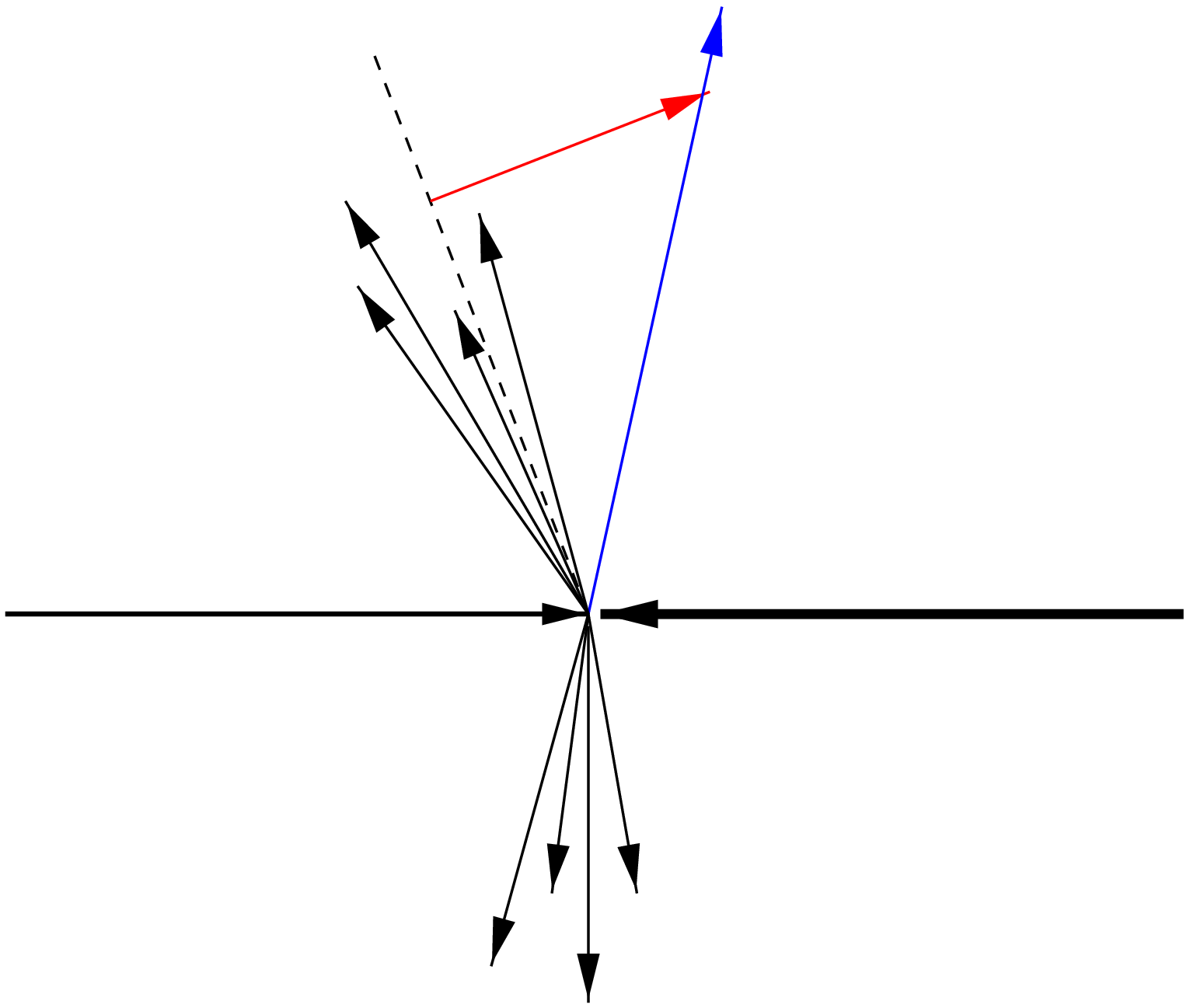}}
\put(-11,-2){\Large\red $\mu$}
  \put(-0.5,3.2){\Large $e$}
  \put(10.4,3.2){\Large $p$} 
  \put(6.5,8.0){\blue{\large $l$}} 
  \put(2.,8.5){\blue{\large\sf jet}} 
  \put(4.0,7.5){\red{\Large $\vec{p_t}^{\mbox{\small rel}}$}} 
\end{picture}
\caption{\label{fig:event-ptrel}
A beauty candidate in the H1 detector; definition of $\ptr$.}
\end{figure}

All HERA measurements of $b$ production 
so far rely on inclusive semi-leptonic decays,
using identified muons or electrons in dijet events. 
A beauty candidate with a distinctive two-jet structure and 
the clean signature of a penetrating muon track is shown 
in figure~\ref{fig:event-ptrel}. 
\unitlength0.54cm
In order to discriminate the $b$ signal from background sources,
two observables have been used which 
are based on the mass of the $b$ quark or on its lifetime. 
The high mass gives rise to large values of the  
transverse momentum $\ptr$ of the decay lepton relative to 
the direction of an associated jet (see figure~\ref{fig:event-ptrel}).
Both collaborations have published 
photoproduction results~\cite{h1openb,zeusopenb} using this method.
More recently, 
with the precision offered by the H1 vertex detector~\cite{cst} 
it has become possible
to observe tracks from secondary $b$ vertices
and to exploit the long lifetime as a $b$ tag.  
  
\begin{figure}[t] 
\unitlength0.48cm
  \begin{picture}(16.0,13.0)(0,.4)
\put(-0.,0.0){\epsfig{file=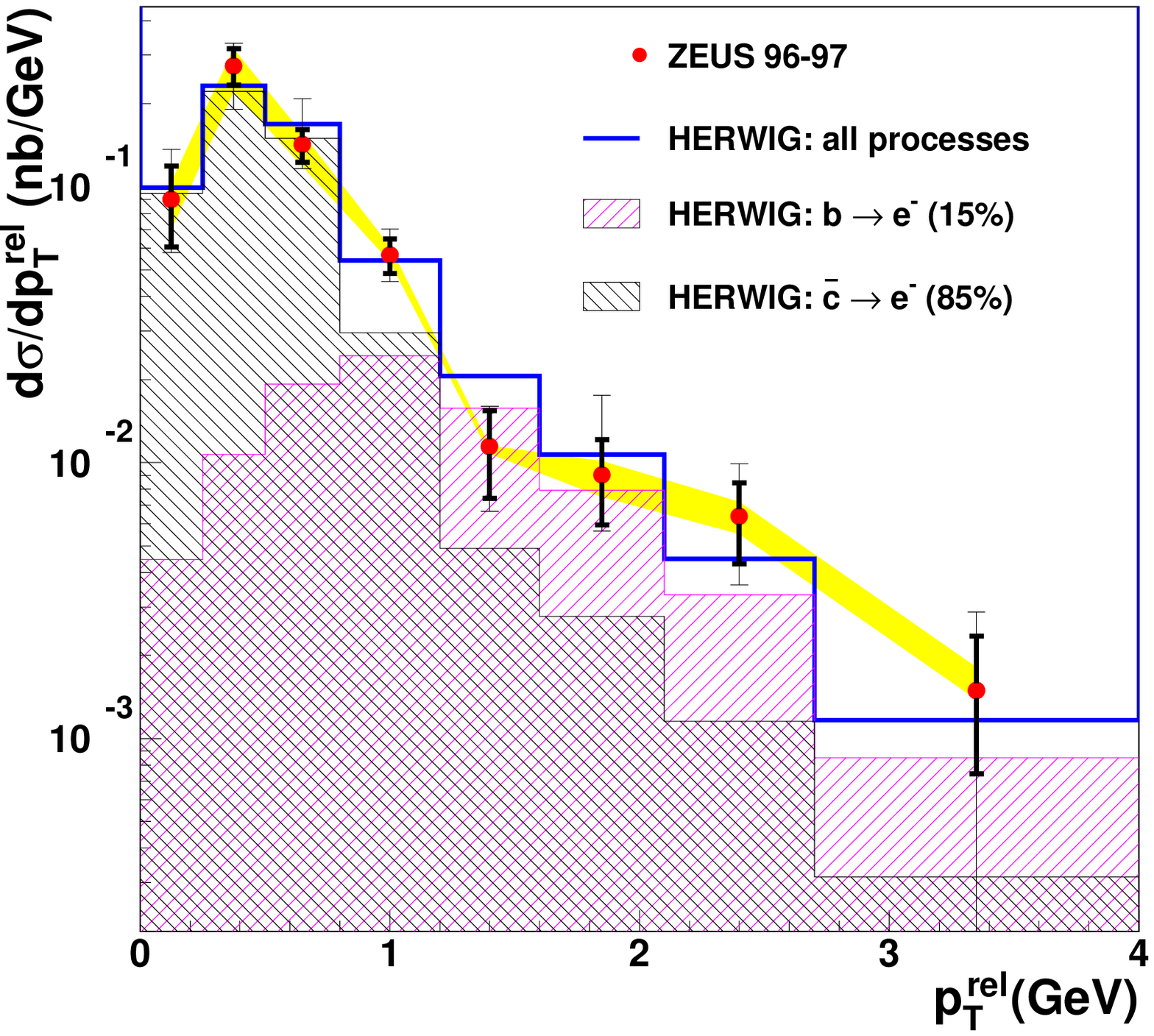,width=7.2cm}}
  \end{picture}
\unitlength0.45cm
  \begin{picture}(16.0,10.0)
\put(-0.6,0.0){\epsfig{file=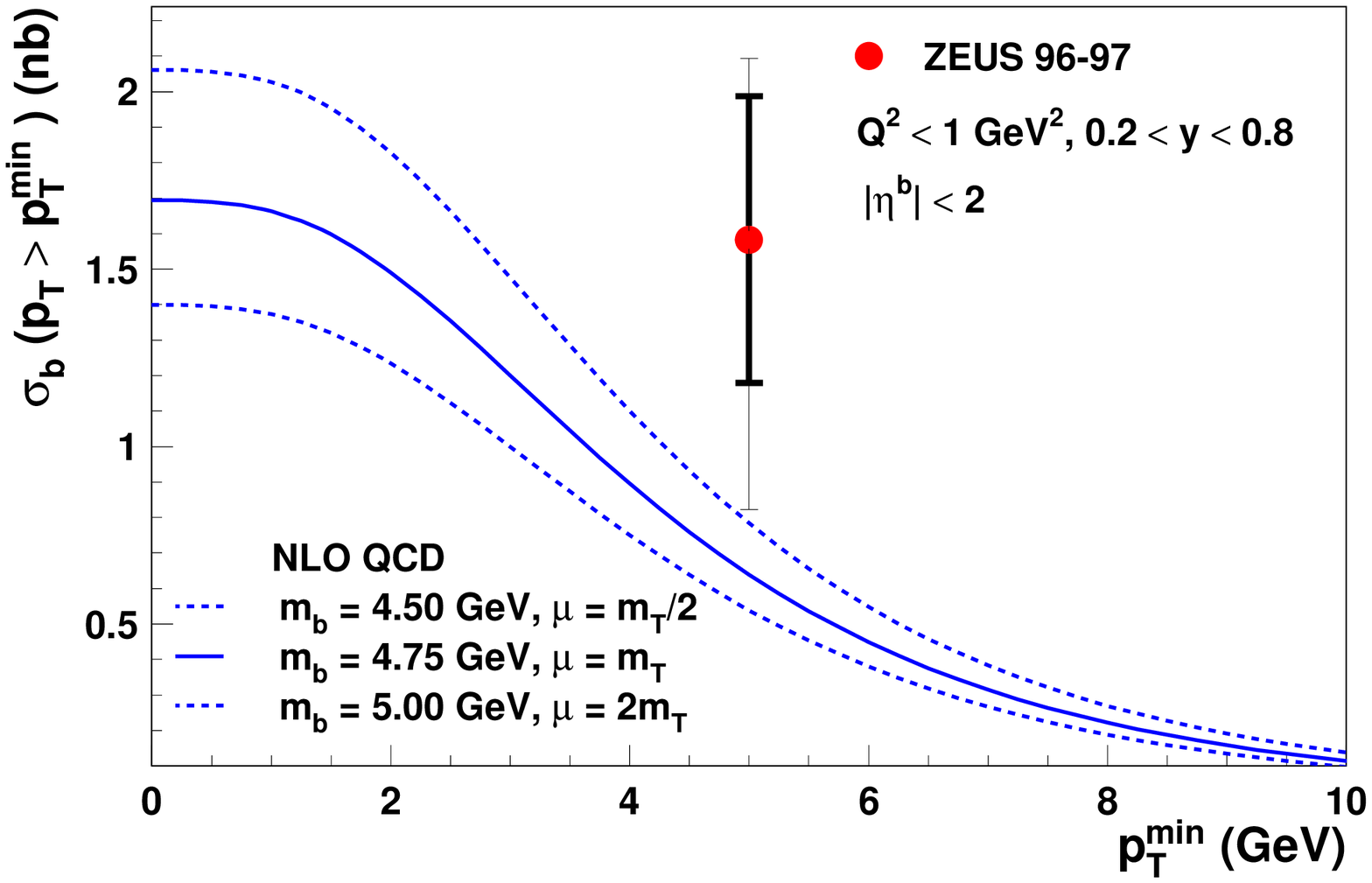,width=8.1cm}}
  \end{picture}
\caption{\label{fig:zeusresult}
Cross section for the production of electrons in 
dijet events, as a function of $\ptr$;
extrapolated $b$ quark cross section
compared with NLO QCD.}
\end{figure}

The ZEUS analysis has been discussed in its preliminary 
version at the previous Ringberg workshop~\cite{pitzl}
and has meanwhile been published~\cite{zeusopenb}.
It uses electrons 
which are identified in the ZEUS detector by means of the topology of 
calorimetric energy deposition and of the specific
energy loss $dE/dx$ measured in the central track detector. 
Background from hadrons misidentified as electrons is statistically 
subtracted. Non-prompt electrons, mostly from photon conversions,
are identified on the basis of track topology and invariant mass criteria. 
From a sample of $e^+p$ data corresponding to an integrated luminosity of 
38.5~\pbmo, a signal of $943\pm 69$ electrons attributed to 
heavy quark decays is extracted. 
Performing the procedure in bins of $\ptr$ and correcting for efficiency, 
the differential cross section shown in figure~\ref{fig:zeusresult} 
is obtained. 
Overlaid is the result of a Monte Carlo simulation of beauty and charm 
production, using the HERWIG generator~\cite{herwig}. 
The normalization is adjusted to that of the data, and the shape
is fitted by varying the relative contributions of charm and beauty.  
The fit yields a beauty fraction of $f_b = (14.7 \pm 3.8)\,\%$ 
in the sample,
close to the HERWIG expectation of 16~\%. 
This is translated into a cross section 
for the visible kinematic range,
\begin{equation}
\sigma^{b \ra e^-}_{e^+ p \ra e^+ + \mbox{\rm  dijet} + e^- + X}        
= 24.9 \pm 6.4 ^{+4.2}_{-7.3} \ {\rm pb}
\end{equation}
The range is 
defined by the requirements on the photon virtuality, $Q^2<1$ GeV$^2$, 
the inelasticity variable $0.2<y<0.8$, 
the jet rapidity $|\eta|<2.4$, the transverse jet energy 
$E_T^{\mbox{\rm jet}} > 7 (6)$ GeV 
for the (second) most energetic jet, 
the transverse momentum $p_T^{\mbox{\rm e}}  > 1.6 $~GeV
and rapidity $|\eta|<1.1$ of the electron.

In a similar manner, the cross section has also been measured as 
a function of $x_\gamma$, an observable which is calculated from the 
energies and rapidities of the jets and which 
in the leading-order picture corresponds 
to the momentum fraction of the parton in the  resolved photon entering 
the hard process. 
(For direct interactions of the photon $x_{\gamma}\approx 1$.)
The $x_{\gamma}$ spectrum allows a resolved contribution of 
$(28\pm 5(stat.))\;\%$
to be estimated in this picture. 
The HERWIG expectation is 35 \%, 
which in this program is mostly due to flavour excitation in the photon. 
Due to the small beauty component in the sample, this does not 
allow conclusions on the $b$ production dynamics to be drawn, but it 
corroborates earlier observations on charm, 
made by ZEUS with $D^{\ast}$ mesons in dijet events~\cite{zeusgpcharm}.  

The visible cross section can directly be compared with leading order 
Monte Carlo predictions, which are 8~pb for HERWIG~\cite{herwig}
and 18~pb for PYTHIA~\cite{pythia} (also including flavour excitation). 
The recently developed CASCADE program~\cite{cascade}, 
based on unintegrated parton distributions and the 
CCFM evolution equation~\cite{ccfm},
follows a different approach to account for higher order QCD effects.  
As shown in a separate contribution to this workshop~\cite{hjungrb}, 
the approach is able to reproduce results on $b$ production at the Tevatron. 
The CASCADE prediction for HERA 
in the ZEUS kinematic range is 18 pb~\cite{hjungpriv}.

In order to compare the measured cross sections with QCD predictions, 
the Monte Carlo simulation is used to convert the result 
into a $b$ quark cross section for a range 
restricted in terms of parton kinematic variables to 
$p_T^{\mbox{\rm b}}  > 5 $ GeV, $|\eta^{\mbox{\rm b}}|<2$
and to the same $Q^2$ and $y$ range as above. The result,  
\begin{equation}
\sigma^{\rm ext}_{e^+p \rightarrow
e^+bX} = 1.6 \pm 0.4 (stat.) ^{+0.3}_{-0.5} (syst.) ^{+0.2}_{-0.4}
(ext.)~\mbox{nb}
\end{equation}
with the third error indicating the model dependence of the extrapolation, 
is shown in figure~\ref{fig:zeusresult} together with the NLO QCD 
expectation in the massive approach, which is obtained with the 
FMNR program~\cite{fmnr}
and drawn as a function of the minimal $p_T^{\mbox{\rm b}}$
requirement.
The theoretical curve lies below the measurement. 

The first observation of $b$ production at HERA by H1~\cite{h1openb} 
was based on the $\ptr$ method, too, and has also been presented
at the previous workshop of this series~\cite{pitzl}.
We focus here on the more recent work based on the lifetime signature, 
which improves the photoproduction result~\cite{h1bosaka} 
and provides a first measurement in DIS~\cite{h1bbudapest}. 

The H1 central silicon tracker (CST)~\cite{cst} 
consists of two cylindrical layers of silicon strip detectors, 
surrounding the beam pipe at radii of $R=57.5$~mm and $R=97$~mm,
respectively, from the beam axis. 
With an effective length of 358~mm it covers 
a large part of the $ep$ interaction region
and has a polar angle acceptance of $30^0 <\theta< 150^0$ for the outer layer,
for particles emanating from the nominal interaction point. 
Double sided silicon detectors with readout strip pitches 
of 50 $\mu$m and 88 $\mu$m provide
resolutions of 12 $\mu$m in $r\phi$ and 25 $\mu$m in $z$.
The analyses presented here are performed in the transverse plane. 
For tracks with CST hits in both layers, 
the achieved resolution of the transverse distance $dca$ 
to the center of the H1 detector can be parameterized as
$\sigma_{dca} \approx 40\;\mu\mbox{m} 
\oplus 100 \;\mu\mbox{m} /p_T [\mbox{GeV}]$. 
The first term represents the intrinsic resolution 
and the second the contribution from multiple scattering in the beam pipe.

\begin{figure}[t]
  \begin{picture}(28,15)(-.3,-2)
\put(0.,-1.7){\epsfig{file=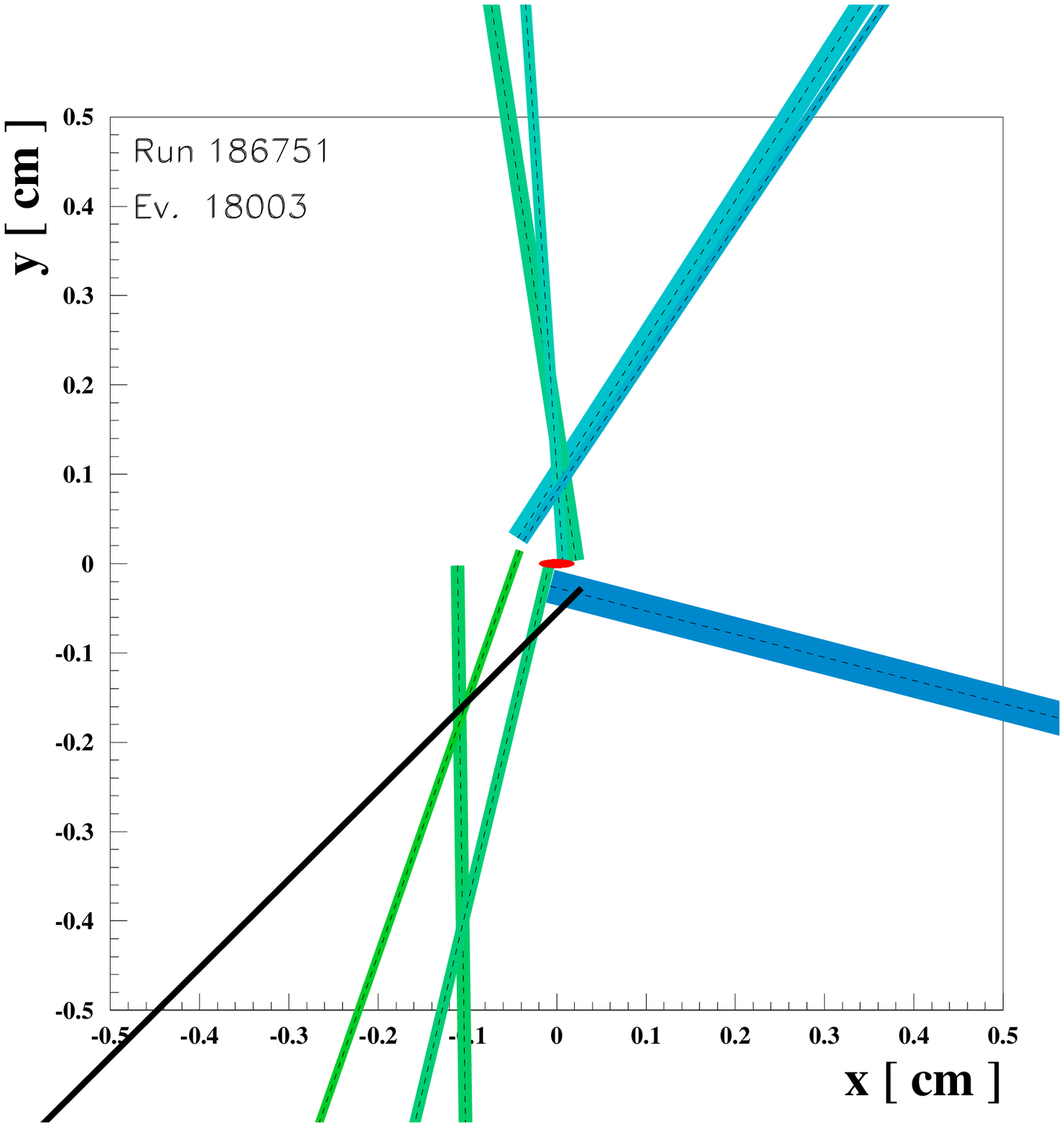,
width=7.8cm}}
\put(2.6,1.5){\Large\red $\mu$}
\put(8.2,6){\normalsize\sf beam spot}
  \unitlength0.66cm
  \put(15.5,8.5){\Large\red\sf{impact parameter}}
  \put(13.0,0.0){\epsfig{angle=0,width=6.6cm,file=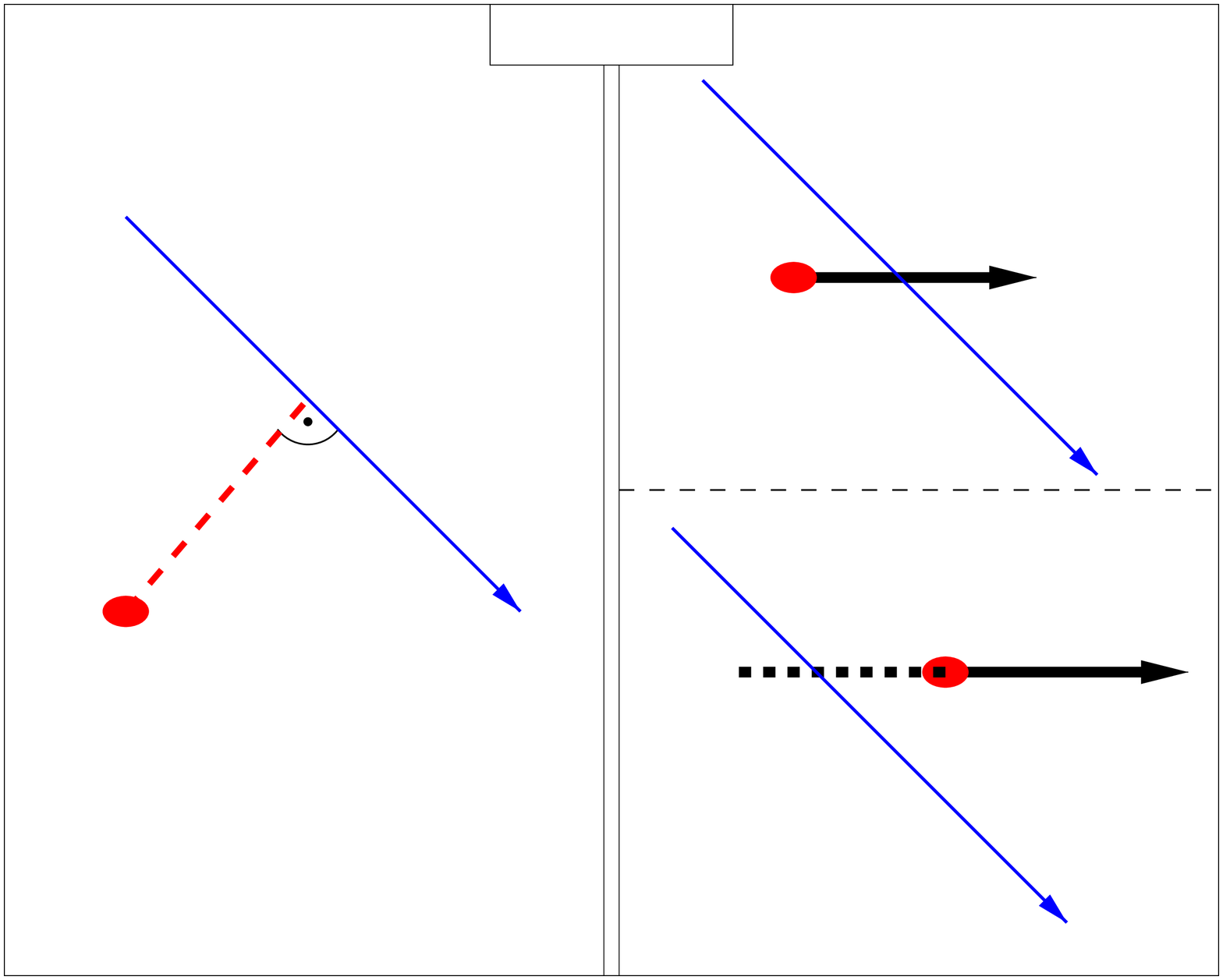}}
  \put(17.2,7.6){\sf\boldmath$\otimes$ beam}
  \put(13.5,2.0){\large\blue\sf primary} 
  \put(13.5,1.3){\large\blue\sf vertex} 
  \put(14.5,6.2){\large\blue\sf lepton} 
  \put(13.5,4.3){\red\huge\boldmath $|\delta|$}
  \put(18.3,4.5){\red\Large\boldmath $\delta>0$}
  \put(18.3,0.5){\red\Large\boldmath $\delta<0$} 
  \put(22.0,5.6){\large\blue\sf Jet} 
  \put(22.0,1.8){\large\blue\sf Jet} 
\end{picture}
\caption[]{\label{fig:vertex-delta}
Vertex region of the event in figure~\ref{fig:event-ptrel}
(view transverse to the beam); definition of the impact parameter $\delta$.} 
\end{figure}

A magnified view of the vertex region of the event of 
figure~\ref{fig:event-ptrel}
is displayed in figure~\ref{fig:vertex-delta}. 
  \unitlength0.6cm
The tracks measured in the CST are represented as bands with widths
corresponding to their $\pm 1\sigma$ precision.
The resolution provided by the CST reveals that
the muon track originates from a secondary vertex well separated from the 
beam spot ellipse. 
A simple and robust technique to exploit such signatures 
is using the impact parameter $\delta$.
Its magnitude is given by the $dca$
of the track to the primary event vertex,
and its sign is positive or negative, 
depending on the intercept of the track with the jet axis 
being downstream or upstream of the primary vertex 
(see figure~\ref{fig:vertex-delta}). 
Decays of long-lived particles are signalled by positive 
impact parameters, whereas the finite track resolution yields 
a symmetric distribution. 

In order to establish the method, the lifetime-based analysis
is performed by applying a similar selection 
of dijet events with identified muons as 
for the published $\ptr$ analysis in photoproduction. 
Jets are reconstructed here using the inclusive $k_t$ algorithm~\cite{kt} 
and required to have transverse energies $E_T>5$ GeV.
For each muon candidate track,
the impact parameter $\delta$ is calculated,
which requires the precise knowledge
of the $ep$ interaction point.
The transverse profile of the interaction region at HERA
has a Gaussian width of about $150 \; \mu\mbox{m}$  in the horizontal
and of about $40\; \mu\mbox{m}$ in the vertical direction.
Average beam coordinates determined from many 
consecutive events are used to constrain 
the primary vertex fit applied to 
CST measured tracks in each individual event,
excluding the muon track under consideration.
A typical $\delta$ resolution
of 90 $\mu$m has been achieved with comparable contributions
from the muon track and from the primary vertex. 

\begin{figure}[t]
\begin{center}
\unitlength6mm
\begin{picture}(18,12)
\put(1., -1.){\epsfig{figure=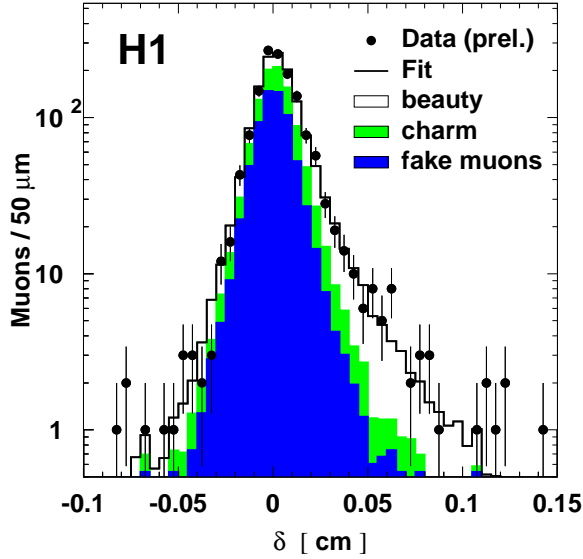,height=9cm}}
\end{picture}
\caption{\label{fig:delta}
Muon impact parameter distribution for the photoproduction sample 
and decomposition from the likelihood fit.}
\end{center}
\end{figure}

DIS and photoproduction events are analyzed separately according to whether
or not the beam positron is scattered into the main detector.  
Shown in figure~\ref{fig:delta}
is the impact parameter distribution for 
the photoproduction data which correspond to an integrated luminosity 
of ${\cal L}=14.7$~pb$^{-1}$ and contain 1403 events with 
$N_{\mu} = 1415$ muon candidates.
The spectrum is decomposed by a maximum likelihood fit 
which adjusts the relative contributions from beauty, charm and
fake muons to the sample.           
The fit describes the data well and
yields a $b$ fraction of $26\pm5\,$\%, which
translates into a visible cross section of 
$\sigma^{vis}_{ep\ra\bbb X\ra \mu X} =159\;\pm\;30 \pm 29\;$~pb 
in the kinematic range defined by 
$p_T(\mu)>2$ GeV,
$35^\circ<\theta(\mu)<130^\circ$ and 
$Q^2<1$ GeV$^2$, $0.1<y<0.8\,$.
Using an independent
signature and new data,
this confirms the published result, 
$\sigma_{ep}^{vis}=176 \pm 16\; ^{+27}_{-17}$~pb 
in the same range~\cite{h1openb}. 
The fitted charm fraction is also compatible
with the H1 measurement of $D^{\ast}$ photoproduction~\cite{h1gpcharm}.

\begin{figure}[t]
\begin{center}
\unitlength8.4mm
\begin{picture}(18,8)(0,0.5)
\put(0, 0.){\epsfig{figure=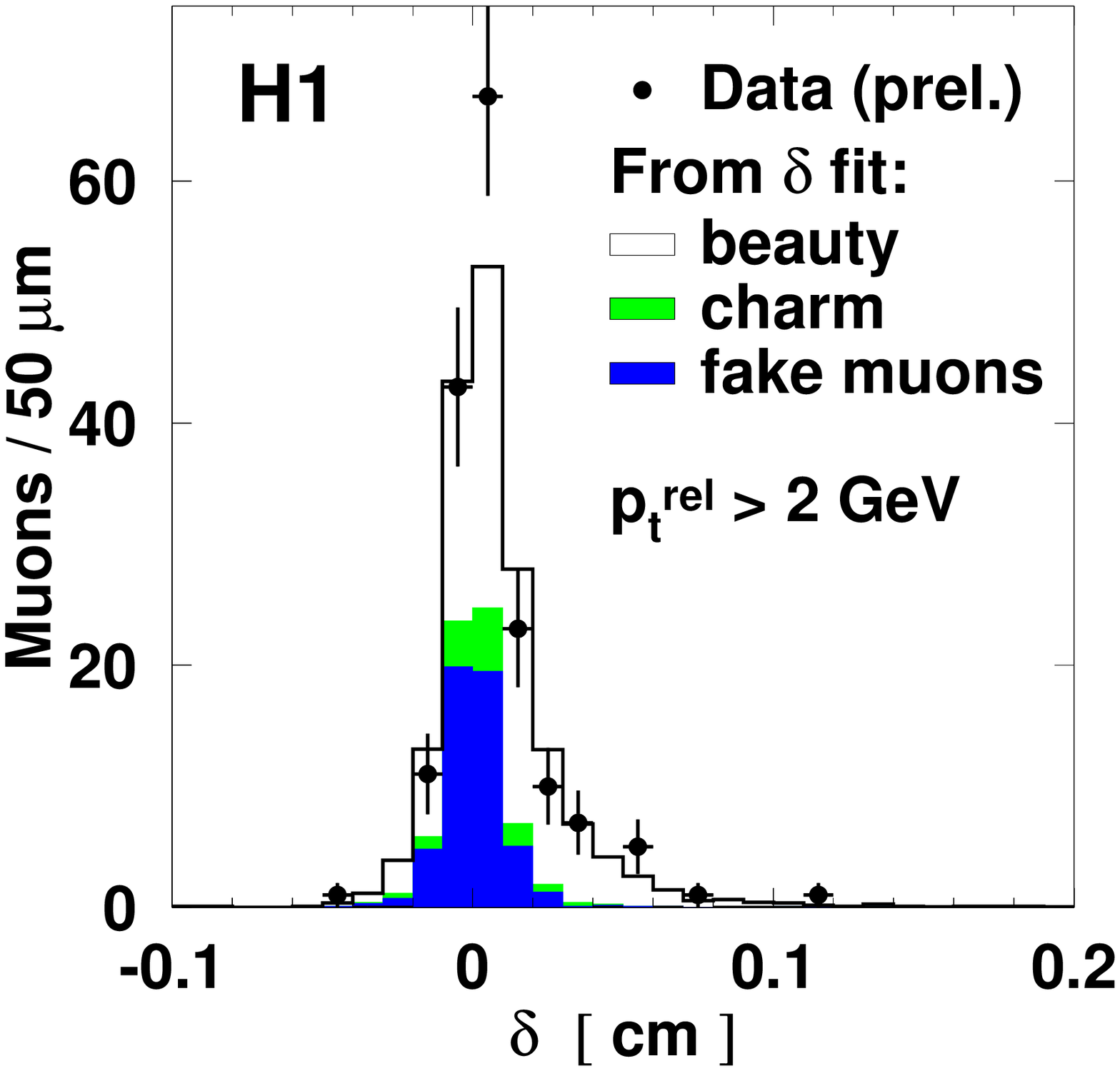,height=8.4cm}}
\put(9.5, 0.){\epsfig{figure=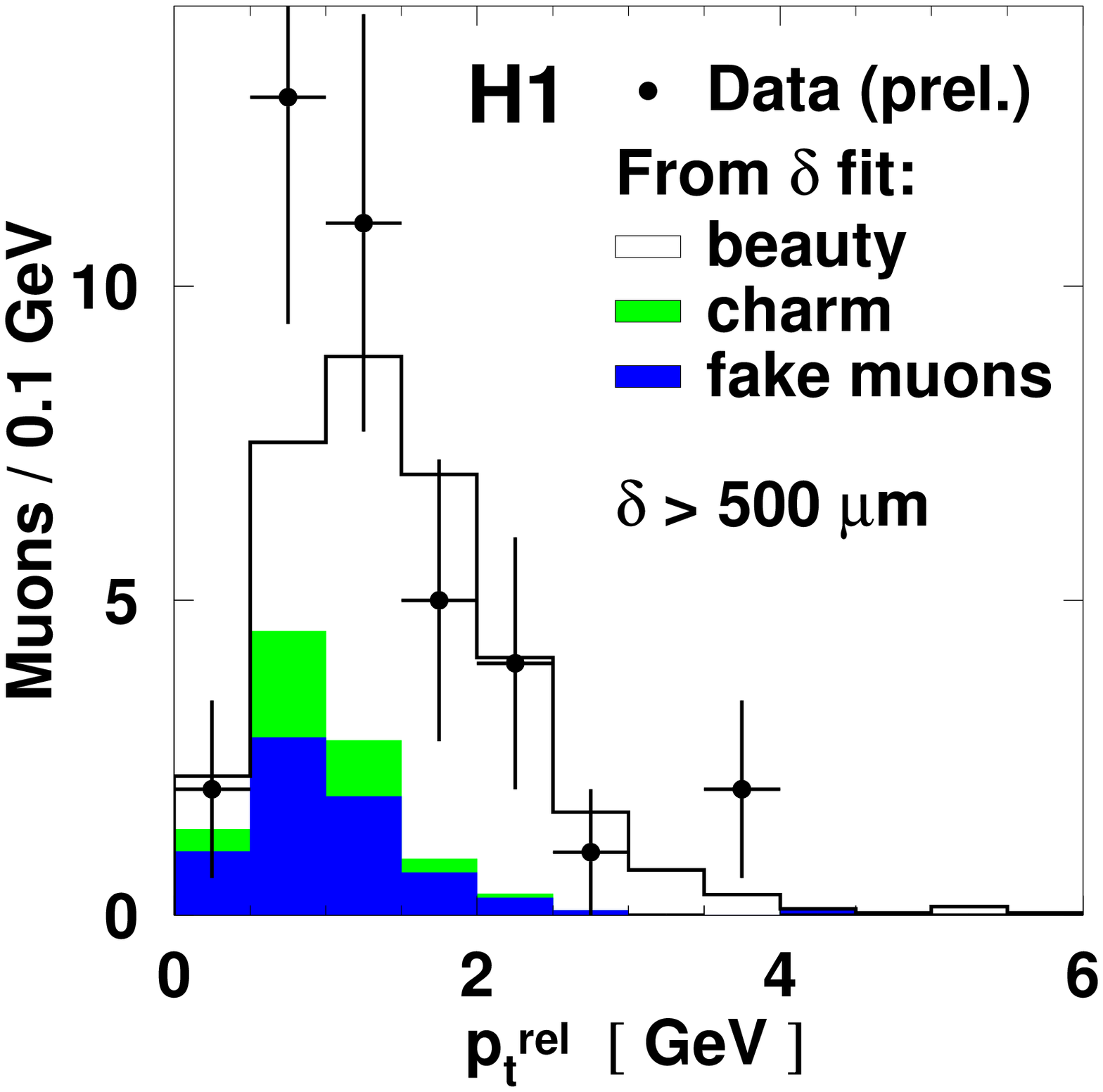,height=8.4cm}}
\end{picture}
\caption{\label{fig:enrich}
Muon impact parameter and $p_T^{rel}$ distributions for 
$b$ enriched photoproduction samples, 
with estimated contributions.}
\end{center}
\end{figure}

To further establish the consistency of the sample composition in the 
two observables
$\delta$ and $p_T^{rel}$, the $b$ component
in the events is enriched 
by restricting the range of one variable and then studying
the distribution of the other.
Figure~\ref{fig:enrich} shows the observed $\delta$ spectrum 
after a $p_T^{rel}$ cut.
The different contributions
shown in Figure~\ref{fig:enrich}
are the absolute predictions for the limited $p_T^{rel}$ region,
evaluated from the $\delta$ fit to the full sample.  
The observed impact parameter spectrum and the fit prediction,
with a dominating beauty component, agree within errors.
{\it Vice versa}, the $p_T^{rel}$ spectrum after a
cut on $\delta$ agrees within errors with the fit prediction 
for a $b$ enriched sample. 
Since the two observables are consistent and only weakly correlated, 
they can be combined
in a likelihood fit to the two-dimensional $(\delta, p_T^{rel})$ distribution.
It yields $f_b = (27 \pm 3)\,\%$ and 
\begin{equation}\label{eq:hg1gp}
\sigma^{vis}_{ep\ra\bbb X\ra \mu X} =160\;\pm\;16 \pm 29\;{\rm pb},
\end{equation}
which is again consistent with the previous H1 result~\cite{h1openb}. 
The average, taking correlated systematic uncertainties into account,
is $\sigma_{ep}^{vis}= 170 \pm 25 \; {\rm pb}\;$.

\begin{figure}[b]
\begin{center}
\unitlength8.4mm
\begin{picture}(18,8)(0.5,0.5)
\put(0,-0.){\epsfig{figure=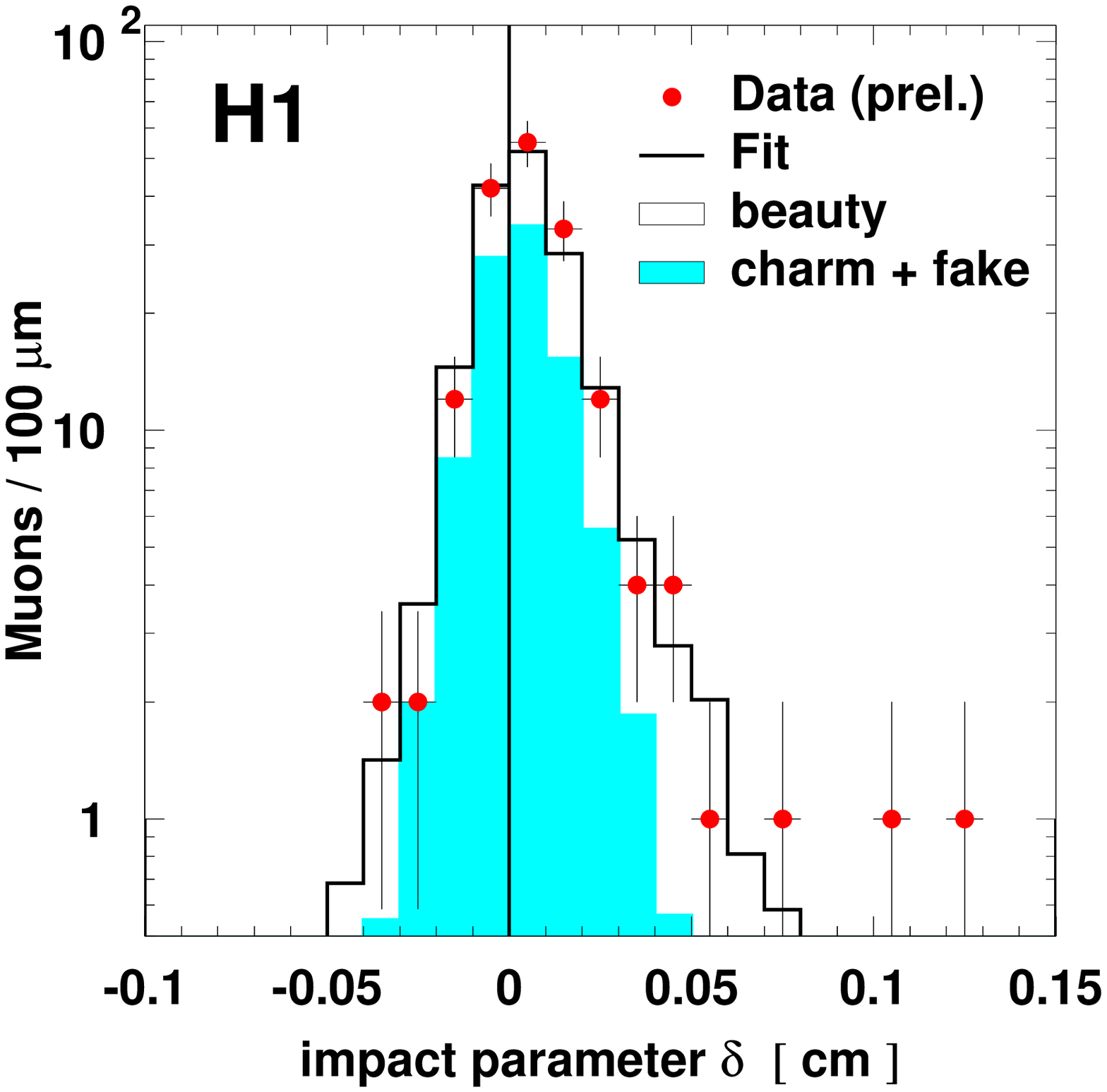,width=8.4cm}}
\put(10.,-0.){\epsfig{figure=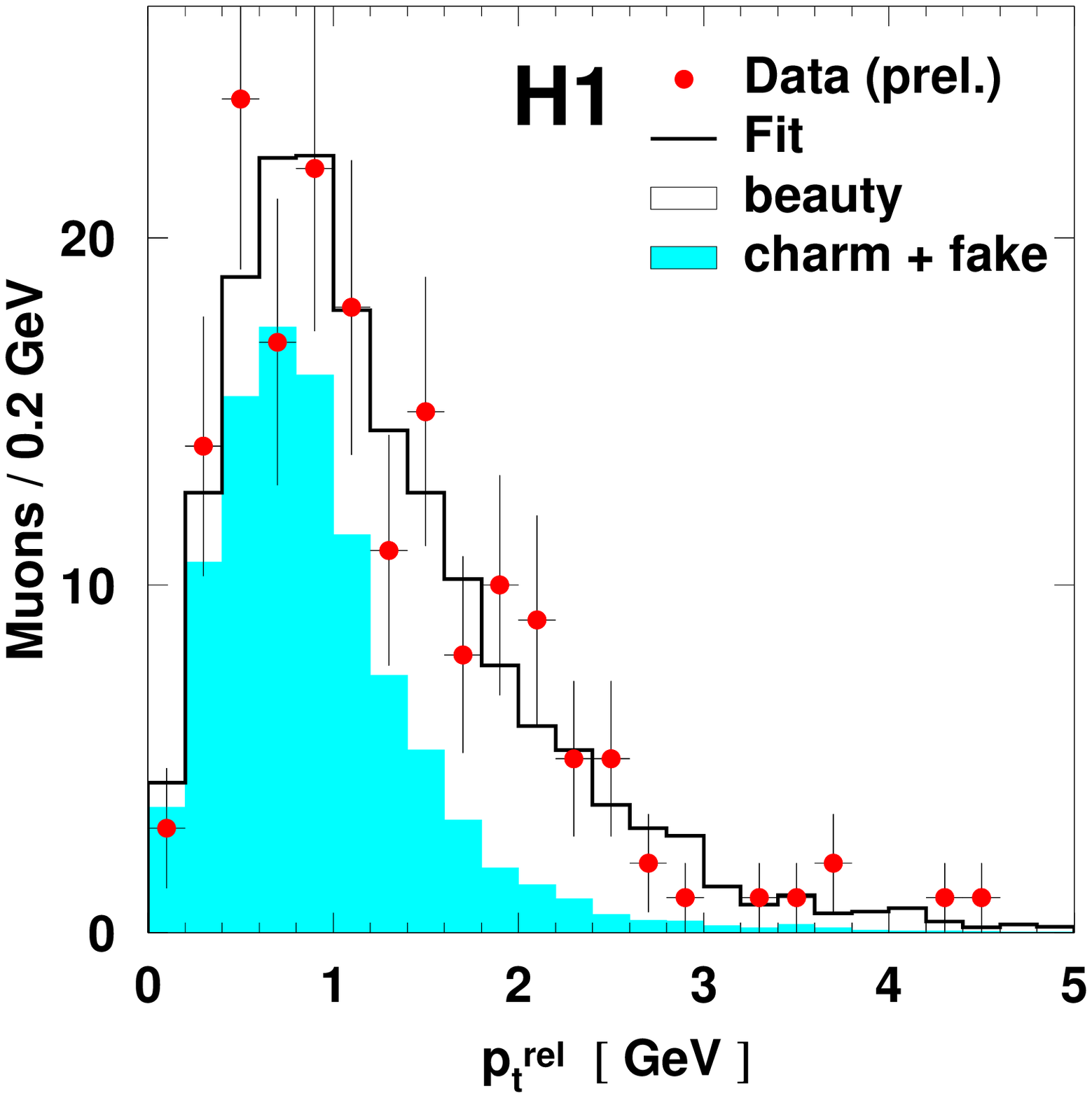,width=8.4cm}}
\end{picture}
\caption{\label{fig:disptdelta}
Muon impact parameter and $p_T^{rel}$ distributions for DIS,
with decomposition from the likelihood fit.}
\end{center}
\end{figure}

The analysis of the smaller DIS sample relies on the sensitivity of the  
combined likelihood fit to the two-dimensional 
distribution in $\delta$ and $p_T^{rel}$.
Using the same jet and muon requirements as in the photoproduction case, 
171 candidates are selected from a dataset corresponding to 10.5~\pbmo.
The projections of the $(\delta,\ptr)$ distribution 
are shown in figure~\ref{fig:disptdelta}
together with the decomposition from the fit 
which yields a $\bbb$ fraction of $f_b = (43\pm 8)\,\%$.
Both variables are well described. The need for a sizable $\bbb$ component 
is evident from the lifetime based signature as well as from the 
$\ptr$ spectrum. 
The fit does not allow to disentangle the background sources themselves 
with meaningful accuracy, but the $b$ fraction is only weakly sensitive 
to the relative amount of charm and fake muons in the background. 
The DIS cross section in the kinematic range 
given by $2<Q^2<100$ GeV$^2$, $0.05<y<0.7\,$, 
$p_T(\mu)>2$ GeV and 
$35^\circ<\theta(\mu)<130^\circ$ is 
\begin{equation}~\label{eq:h1dis}
\sigma_{ep\ra\bbb X\ra \mu X}^{vis} = 
\;39\;\pm\;8\;(stat.)\; \pm 10\;(syst.)\;\;{\rm pb}\ . 
\end{equation}

The visible cross sections can be directly compared to 
NLO QCD calculations using 
the FMNR~\cite{fmnr} and HVQDIS~\cite{hvqdis} programs 
which provide the option to scale the $b$ quark momenta 
with a Peterson fragmentation function~\cite{peterson}
in order to obtain $b$ hadron momenta. 
The distributions were folded with a lepton spectrum 
for $b$ decays extracted from the AROMA~\cite{aroma} Monte Carlo generator. 
The results are $54\pm 9$~pb for photoproduction and 
$11\pm 2$~pb for DIS,  where the error is predominantly due 
to the $b$ quark mass uncertainty. 
The expectations are much lower than the H1 measurements. 
The data have also been compared with the LO Monte Carlo predictions.
The AROMA program~\cite{aroma}, following a ``massive'' approach, 
gives 38 and 9~pb for photoproduction and DIS, respectively. 
The CASCADE Monte Carlo results~\cite{cascade,hjungpriv}
of 66~pb and 15~pb, respectively, also fall considerably 
below the measurements.

We note that both the H1 and ZEUS experiments 
observe only a tiny fraction of the total phase space. 
In photoproduction, the seen numbers of events correspond to 25~pb (H1) 
and 4~pb (ZEUS). 
This difference is partially due to different jet energy and $y$ ranges
and to the fact that ZEUS uses only $e^-$, but H1 muons of both signs 
and in addition includes a contribution of about 15\% of secondary muons.   
Both experiments extrapolate by comparable factors%
\footnote{Note that the ZEUS cross section is corrected for the 
semileptonic branching ratio, while the H1 cross section is not.} 
to compare with theory, but in different ways. ZEUS corrects for the 
lepton selection and translates the jet cuts into quark kinematics,
whereas H1 keeps the lepton cuts but corrects for the jet requirements 
altogether.    
The measurements can best be compared   
when normalized to the same theory. 
However, the comparisons of the experimental results with calculations 
and with each other are still affected by the uncertainties related to the
extrapolations. That these uncertainties are sufficiently assessed by 
using the various available models, remains an assumption not yet backed 
by experimental data. To reduce them, 
the distributions and correlations characterizing the final state 
topologies must be constrained by more refined measurements. 
   
\begin{figure}[t]
\begin{center}
\unitlength8.4mm
\begin{picture}(18,9)
\put(4.,-0.){\epsfig{figure=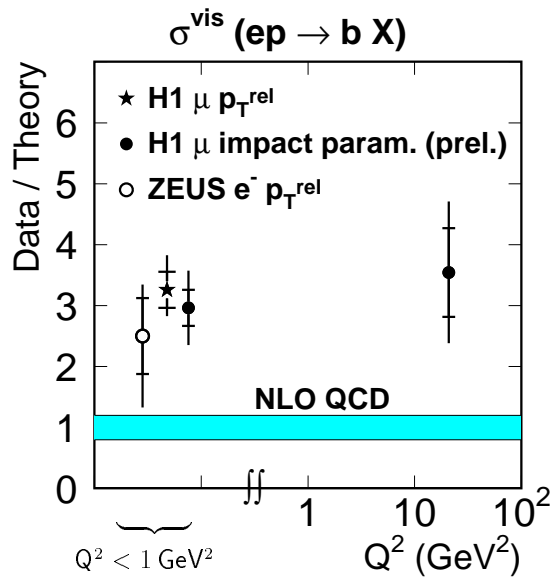,height=7.8cm}}
\end{picture}
\caption{\label{fig:herab}
Ratio of measured $b$ production cross sections at HERA 
over theoretical expectation, as a function of $Q^2$.}
\end{center}
\end{figure}

We summarize the HERA results 
as a function of $Q^2$ in Fig.~\ref{fig:herab}. 
Displayed is the ratio of the measured cross sections over 
theoretical expectations based on the NLO QCD calculations~\cite{fmnr,hvqdis}.
The ratio is consistent with being independent of $Q^2$; 
which indicates that 
the discrepancy between data and theory is not a feature 
of the photoproduction regime alone. 
At larger $Q^2$, 
resolved contributions involving the partonic structure of the photon
are expected to be suppressed~\cite{grs},
the DIS case is therefore complementary and theoretically cleaner.

\section{Beauty production in $\epem$, $\gaga$ and $\pbp$ interactions}

\begin{figure}[t]
  \begin{picture}(26,10)
 \put(0,-0.2){\epsfig{file=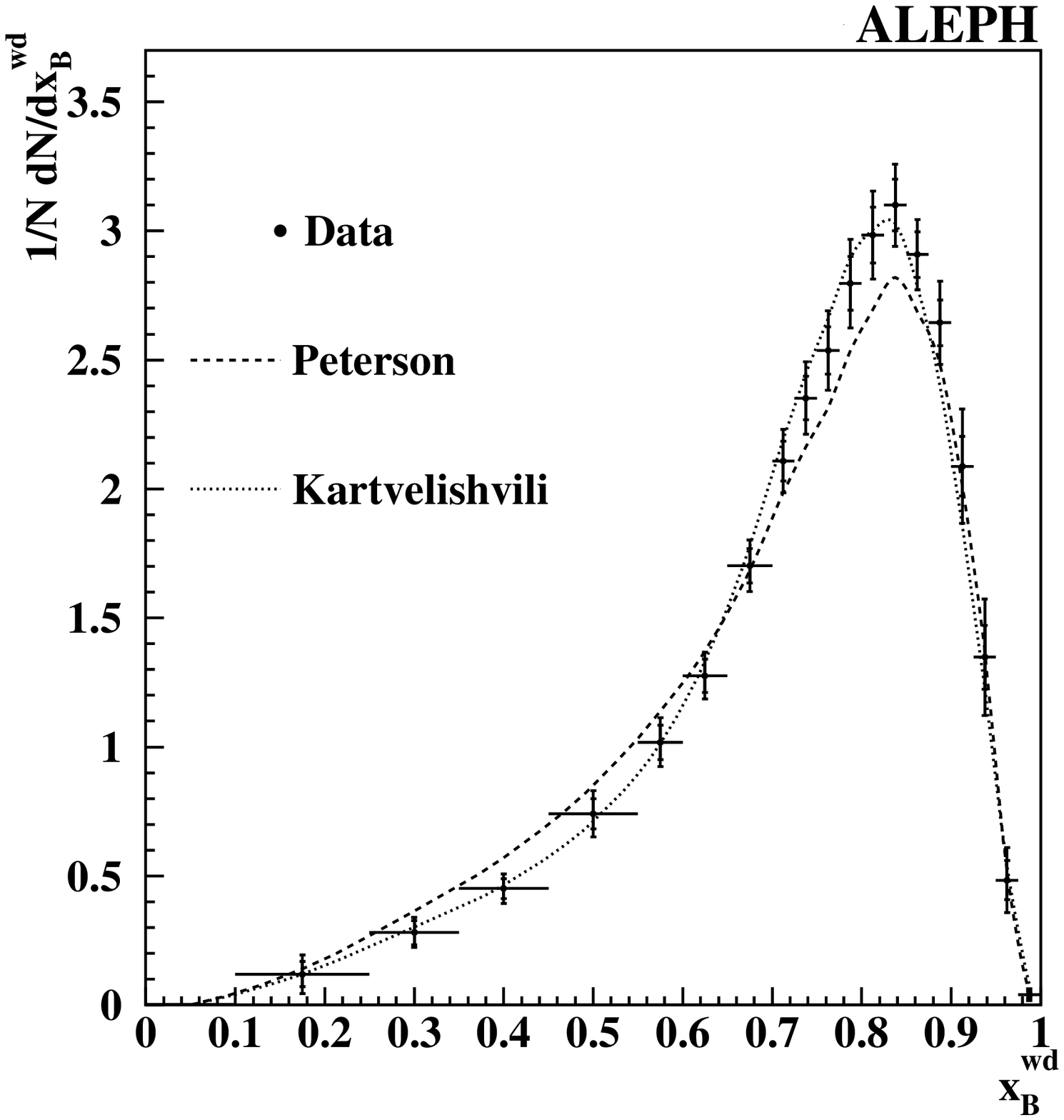,width=7.2cm}}
 \put(11.,-0.2){\epsfig{file=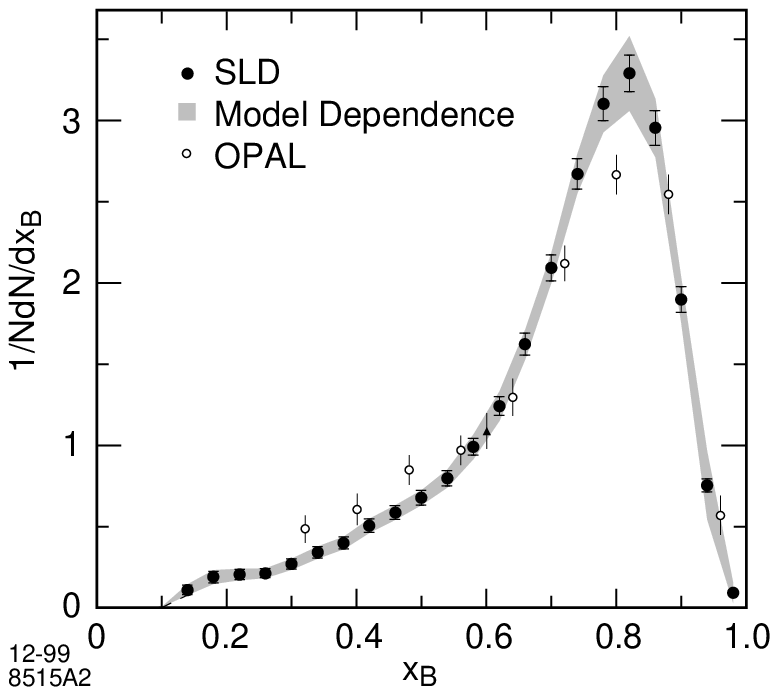,width=7.6cm}}
 \end{picture}
\caption{\label{fig:bfrag}
Measurements of the $b$ hadron fragmentation function.}
\end{figure}

The theoretical understanding of $b$ production data must always  
revert to a description of the fragmentation process, since only hadrons 
can be observed experimentally. 
Since the formation of hadrons involves non-perturbative effects of 
long-range binding forces, it cannot be calculated from first principles. 
Yet, $\epem$ collisions provide a clean laboratory to study 
the process directly and to extract the non-perturbative parameters of the 
fragmentation function. 
This is nowadays best done using data taken at the $Z$ resonance, where 
millions of $\epem\ra Z\ra\bbb$ events have been recorded. 
One observes the distribution of $x_B=E_{wd}/E_{beam}$, the fractional 
energy of the weakly decaying $b$ hadron, normalized to the maximally 
available energy. 
Final data from LEP and SLC with unprecedented precision are now becoming 
available. 
Figure~\ref{fig:bfrag} shows results 
published by ALEPH~\cite{alephbfrag}, 
using $D^{(\ast)}$ meson lepton correlations, 
and by SLD~\cite{sldbfrag}, using an inclusive secondary vertexing technique.
\unitlength7.2mm
Earlier OPAL results~\cite{opalbfrag} with larger errors are shown 
for comparison;
preliminary data from SLD, with even higher statistical precision,
have also been released~\cite{sldbfragprel}.
The ALEPH data are compared to two Monte Carlo simulations, 
which are using different functional forms~\cite{peterson,kartvelishvili}
for the parameterization of the non-perturbative fragmentation function. 
The Peterson form does not describe the data well; a similar 
observation is made by SLD~\cite{sldbfrag}. 

Several comments are in order here. 
The observed $x_B$ spectra reflect the effects of perturbative gluon radiation 
and of the non-perturbative hadronization phase together; 
the models must include both. 
Conclusions from the comparison with data can only be drawn 
for the convolution of the two effects, 
which is here the combination of the Peterson function 
with the leading log parton shower approach~\cite{jetset} 
used to model the perturbative part in the Monte Carlo simulation. 
When combined with resummed NLO calculations~\cite{kniehlhera,olearimerged},
the Peterson function was found to be adequate. 
However, the precision of the new data presents a challenge which 
these calculations have not yet met.

For the interpretation of the HERA measurements,
these discrepancies are too small to be relevant. 
Altogether, the fragmentation results leave little room to modify
QCD predictions for other production modes;
in particular the function cannot be much harder than commonly assumed, 
in contrast to what would be needed to obtain larger predictions 
in the case of cuts on falling $p_T$ spectra.
Since the fragmentation is harder than in the 
charm case, the influence of the non-perturbative {\it versus} perturbative
effects is generally suppressed. 
Once higher accuracy is required, one may have to turn back to the problem 
of how to extract a fragmentation function in a scheme consistent 
with the massive approach. 
While fixed order calculations
could still provide 
an acceptable description of ARGUS data~\cite{argus} on $D^{\ast}$ production,
this appears to be difficult in the case of $B$ mesons produced 
on the $Z$ resonance~\cite{olearieps,olearimerged}.  
 
\begin{figure}[t]
  \begin{picture}(26,12)(-1,0)
 \put(-1,-0.5){\epsfig{file=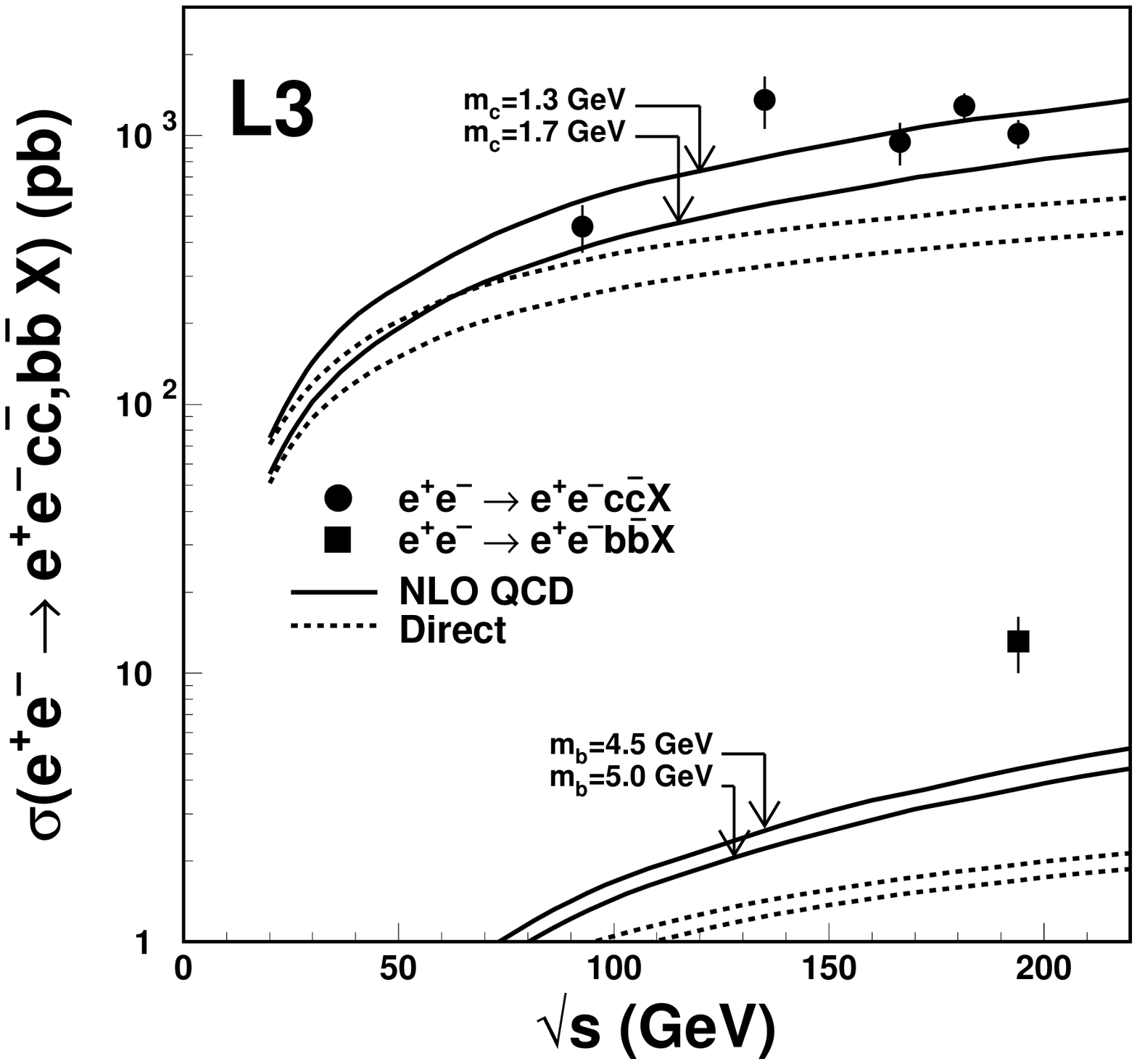,width=7.8cm}}
 \put(13,-0.){\epsfig{file=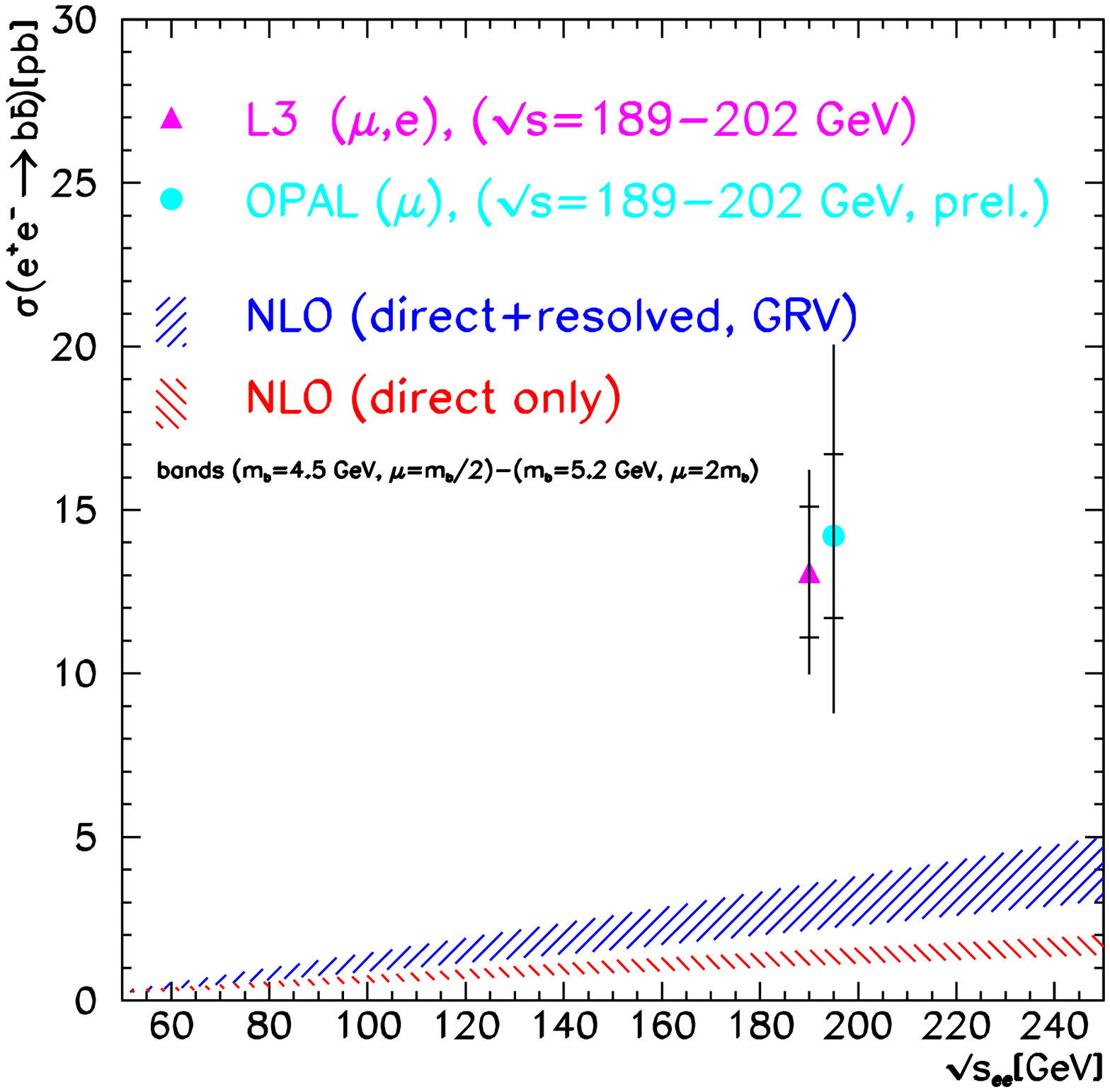,width=7.2cm}}
\put(9,8){\Large\sf\blue charm}
\put(9,5){\Large\sf\red beauty}
\end{picture}
\caption{\label{fig:gaga}
Measurements of charm and beauty production in two-photon collisions.}
\end{figure}

With the increase of the LEP energy towards the 200 GeV range,
it has become possible to observe beauty production in two-photon collisions.
The process is dominated by direct and single resolved interactions,
i.e.\ by photon-photon and photon-gluon fusion, 
which contribute with roughly equal strength~\cite{dreesgaga}. 
The $b$ quarks are predominantly produced at small $p_T$.
They are tagged via identified electrons or muons 
from semileptonic decays, using the $\ptr$ signature 
together with jets with as little as 3~GeV energy.  
The cross section, measured by L3~\cite{l3gaga}, 
is shown in figure~\ref{fig:gaga} together with results for charm production. 
\unitlength6mm
The NLO QCD calculations~\cite{dreesgaga}, 
shown for comparison, have been obtained in the massive approach;
corrections beyond LO amount to about 30~\%. 
While the agreement is good for charm production, the beauty cross section
is underestimated by more than a factor of 2. 
The L3 measurement has recently been confirmed by a preliminary 
OPAL result~\cite{opalgaga}, see figure~\ref{fig:gaga}. 

That the pattern of results resembles the one encountered at HERA
may be not accidental. 
There are many theoretical and experimental similarities,
e.g.\  the necessity to model 
heavy quark fragmentation and jets near threshold.
The QCD approach follows the same scheme;
like in photoproduction it relies on phenomenological input 
to describe the hadronic structure of the photon. 

\begin{figure}[t]
  \begin{picture}(26,11.5)(-1,0)
 \put(-1,-0.5){\epsfig{file=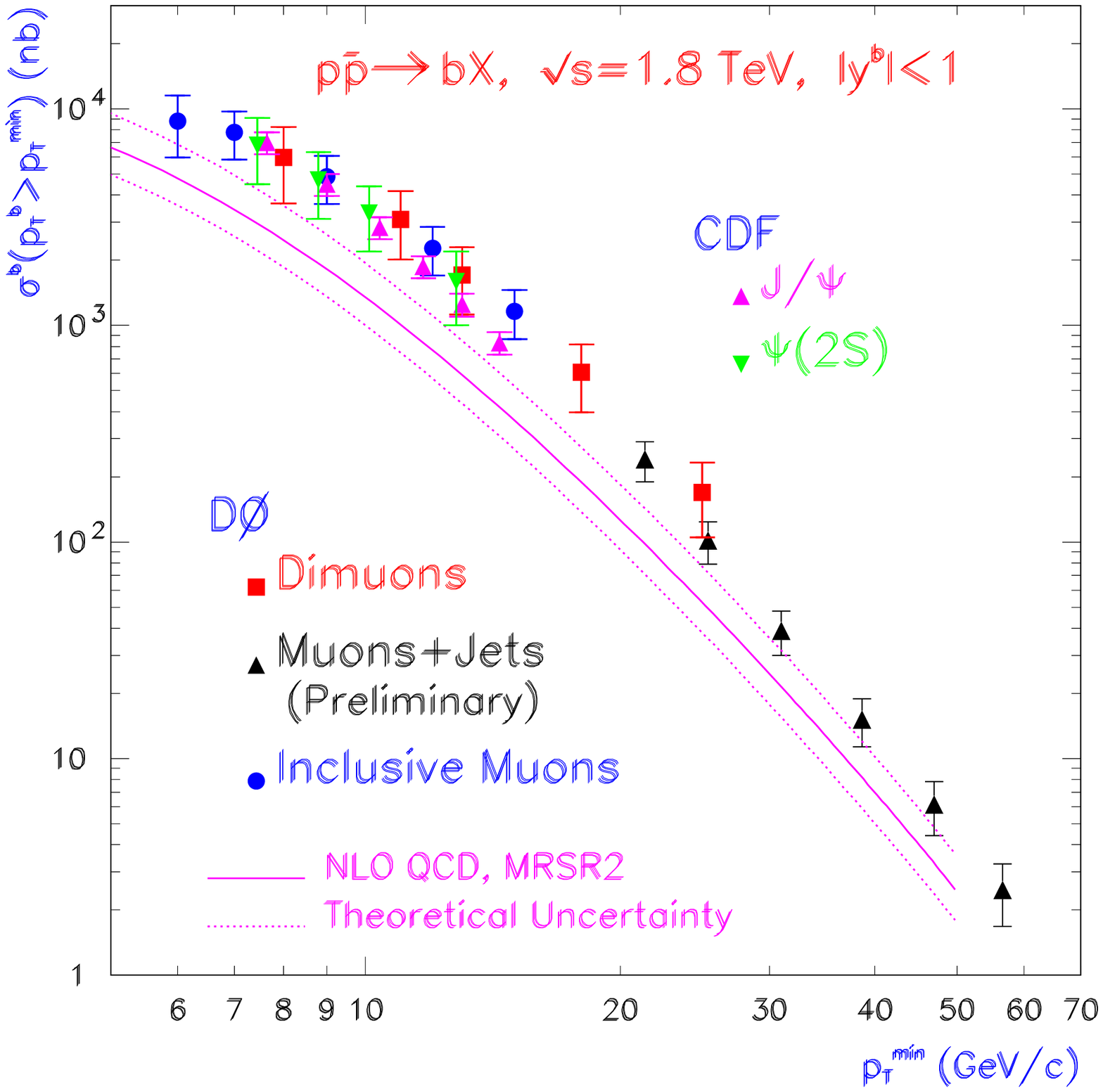,width=7.8cm}}
 \put(13,-0.5){\epsfig{file=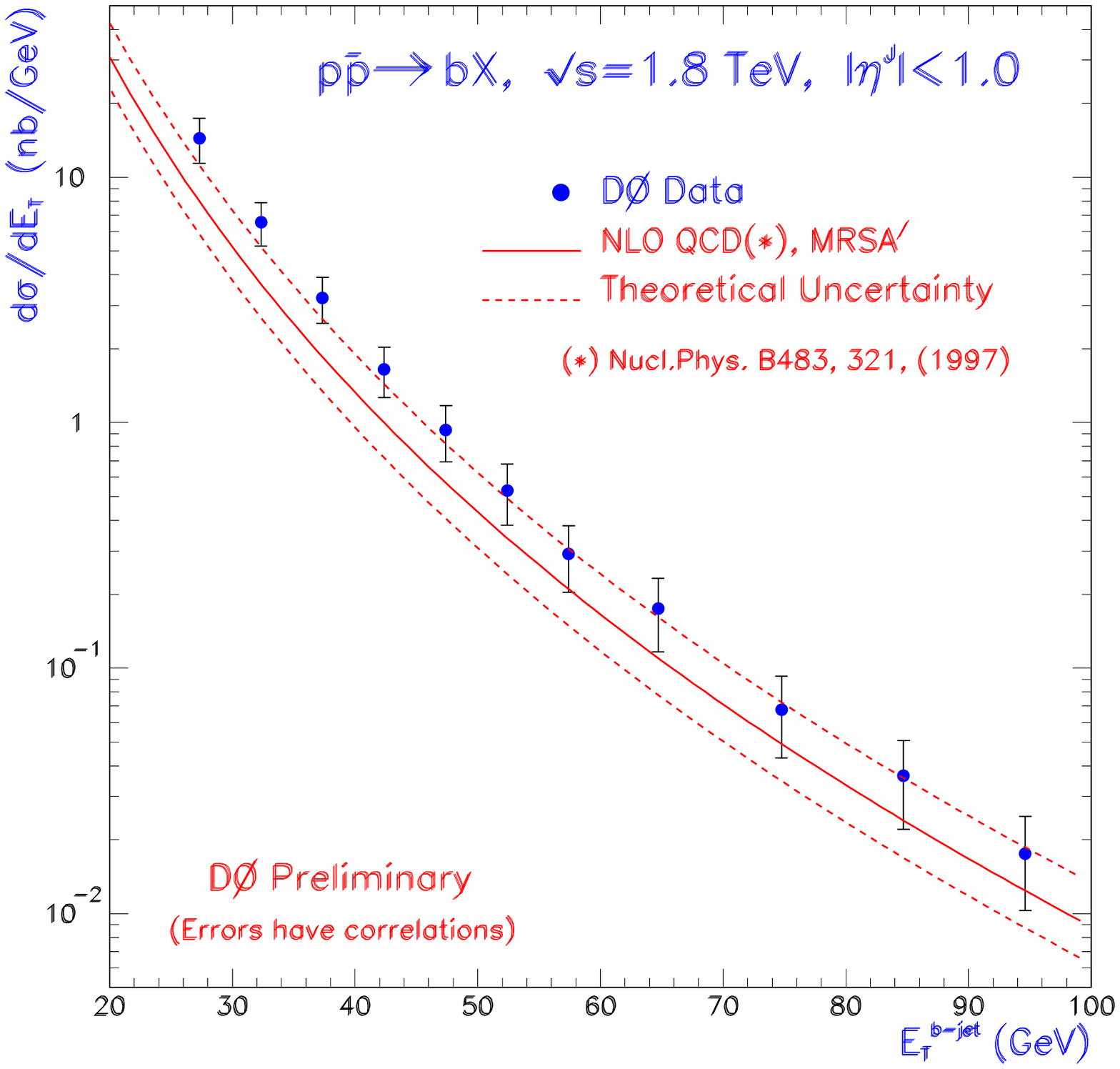,width=7.8cm}}
\end{picture}
\caption{\label{fig:ppxsect}
Measurements of the $b$ cross section, as a function of 
the minimum quark $p_T$;
$b$ jet cross section, as a function of the jet transverse energy.}
\end{figure}

The $b$ production cross section in $\pbp$ collisions has been found in 
excess of QCD predictions
since the turn-on of the Tevatron collider~\cite{bcdfd0}.
At $\sqrt{s}=1.8$~TeV gluon gluon fusion is the dominant mechanism. 
The measurements have meanwhile been performed using a variety of channels 
and techniques, and they cover a large range in transverse momentum.
Results from CDF and D0 are summarized in figure~\ref{fig:ppxsect}. 
The two experiments are consistent with each other.
There are still new results being released, for example using 
signals of exclusively reconstructed $B^+$ mesons~\cite{cdfbplus}, 
which by virtue of secondary vertex detection reside on low background 
and allow measurements with very small systematic uncertainty.  
In general, the cross section is found to be 
above the NLO QCD expectation~\cite{ppmassive}.
However, the figure shows that the excess cannot be absorbed in a 
simple scaling factor, but has some $p_T$ dependence.
We also note that an increase of the excess
with rapidity is observed~\cite{d0brap}.  
While these results are based on $b$ hadron detection, 
it has been suggested to measure 
the $b$ tagged jet cross section instead~\cite{frixijet}. 
The theoretical calculation should be 
safer against soft and collinear effects, and the confrontation
with data less sensitive to assumptions on the fragmentation process. 
The preliminary jet data from D0~\cite{d0bjet}
are also shown in figure~\ref{fig:ppxsect}. 
Indeed, the agreement with theory is better.   
Yet, there is a similar trend for the data to be higher than the central 
prediction, 
so that no disagreement between the two sets of results can be claimed.
\unitlength6mm

\begin{figure}[b]
  \begin{picture}(26,13)(-1,0)
 \put(-1,-0.5){\epsfig{file=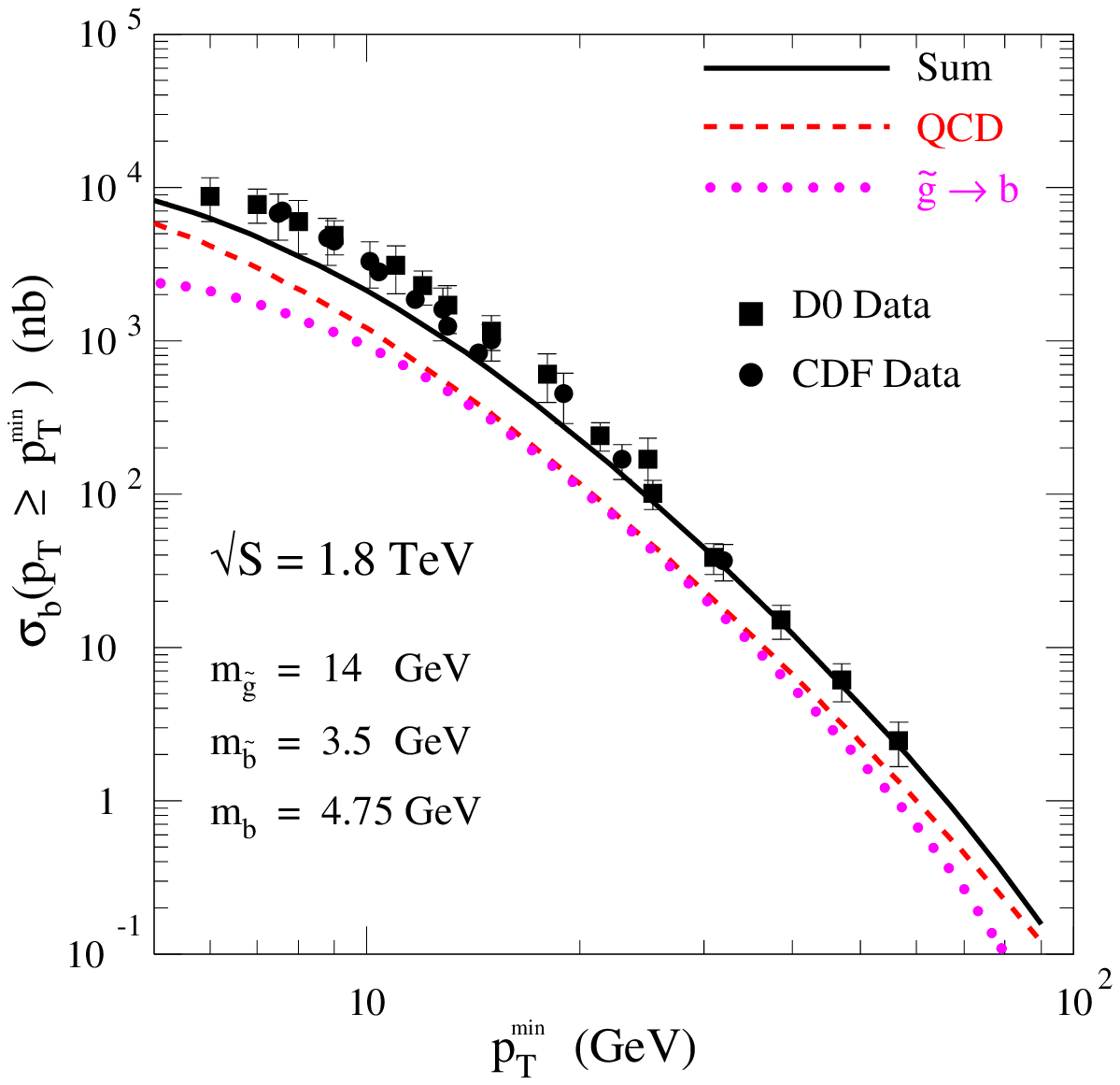,width=7.8cm}}
 \put(13,-0.5){\epsfig{file=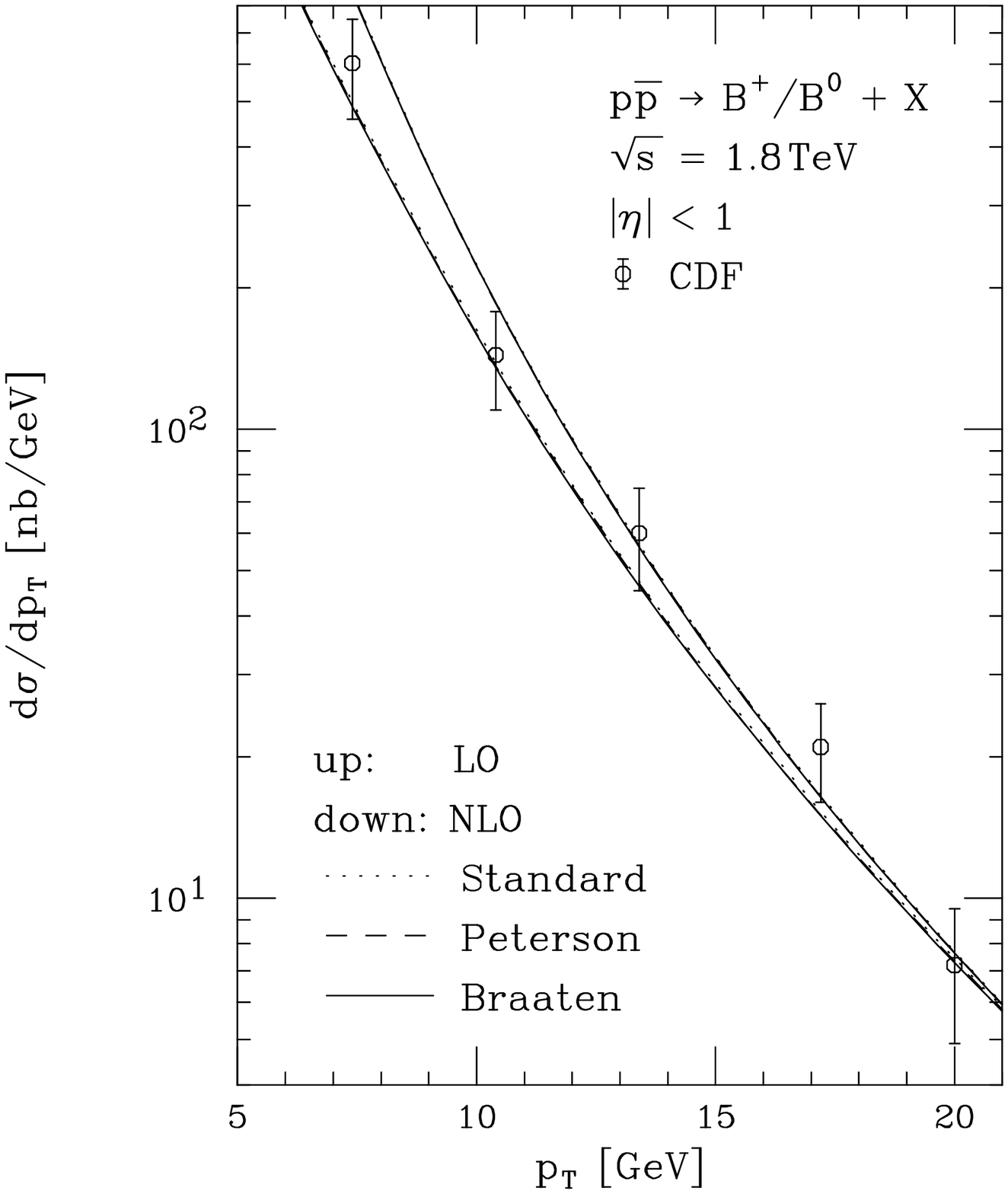,width=7.2cm}}
\end{picture}
\caption{\label{fig:ppxalt}
$b$ cross section data, 
compared with a model assuming light supersymmetric sbottom quark production;
CDF data in comparison with a resummed NLO QCD calculation in the 
massless scheme.}
\end{figure}

The longstanding discrepancy between $b$ hadroproduction data and theory
has stimulated speculations about possible explanations beyond 
the standard model. 
For example, a scenario~\cite{susyb} 
in the context of minimal supersymmetry,
with relatively light gluinos $\tilde{g}$ and sbottom quarks $\tilde{b}$,
can easily reproduce the data, as demonstrated in figure~\ref{fig:ppxalt}. 
\unitlength6mm
There are further consequences that can be tested at the Tevatron; 
for HERA however, first estimates indicate only small effects in this 
particular scenario. 
It is nevertheless interesting 
to note that such an interpretation is not ruled out 
by existing experimental constraints, including the $\epem$ continuum data
and precision measurements or direct searches performed at the $Z$ and 
$\Upsilon (4S)$ resonances
(see~\cite{susyb} and references therein). 

A more conventional approach is to refine the QCD calculations. 
The large scale dependence, which is still present at NLO, 
indicates that important contributions are missing in the 
perturbative expansion of the cross section. 
In figure~\ref{fig:ppxalt}, the $B$ meson cross section~\cite{cdfforkniehl}
is compared to a resummed calculation~\cite{bkniehl} in the massless 
scheme, using perturbative 
fragmentation functions directly adjusted to LEP data. 
A good description is found, surprisingly also for the data at lower $p_T$, 
where the agreement for this scheme is considered as fortuitous 
by the authors of~\cite{bkniehl}. 
However, a similar ``surprise'' was found in the case of 
charm photoproduction at HERA~\cite{h1glue,zeusgpcharm}. 
A cross section enhancement in the medium $p_T$ range of the Tevatron data is 
also found, using high $p_T$ resummation via fragmentation functions in 
a different scheme~\cite{cacciaripp}. 
There are other resummation strategies being pursued~\cite{cascade,ppkt}, 
namely using the concept of unintegrated parton densities, 
which is discussed in more detail in~\cite{hjungrb}. 
These approaches lead to enhancements of the predictions.

The experimental precision can be expected to improve even further, 
when results from the upgraded Tevatron become available, 
and this should provide additional guidelines to find out which 
strategy most effectively includes the missing higher oder contributions. 
In either case, whether speculations on supersymmetry will further be 
nourished, 
or whether ``only'' the QCD techniques will be refined, 
complementary information from HERA will be extremely valuable.
Apart from $\epem$ annihilation, 
DIS represents the cleanest $b$ production environment.

\section{Prospects for the HERA upgrade}

The HERA collider is currently resuming its operation, after an upgrade 
which lays the foundations for an integrated luminosity of  1~fb$^{-1}$ 
to be accumulated in the next 5~years. 
Much of the experimental upgrade program which was pursued in parallel
by the H1~\cite{h1upgrade} and ZEUS~\cite{zeusupgrade}
collaborations is directed towards augmenting the 
capabilities for heavy quark physics, 
notably by improving the acceptance and precision of the tracking systems, 
in particular for forward-going particles, and by enhancing the 
trigger sensitivity for final states with small $p_T$. 

\begin{figure}[htb]
\begin{center}
\unitlength8.4mm
\begin{picture}(18,8.3)
\put(3.,-0.8){\epsfig{figure=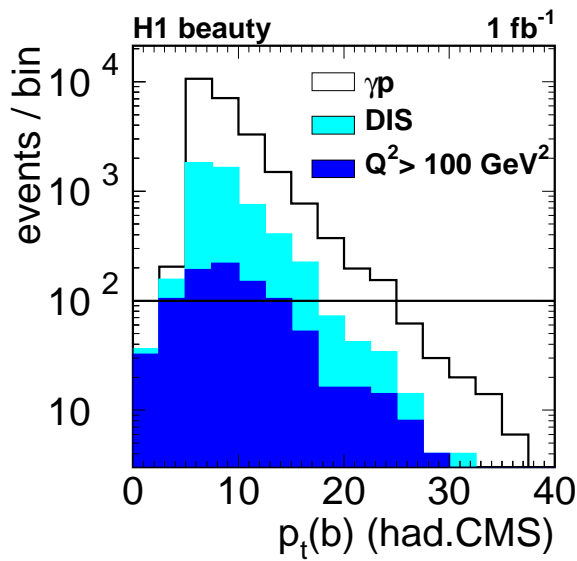,height=8.4cm}}
\end{picture}
\caption{\label{fig:heratwo}
Estimated numbers of events, as a function of the $b$ quark 
transverse momentum in the partonic CMS, for an extension 
of the present H1 muon analysis to the expected full HERA data set. 
}
\end{center}
\end{figure}

The HERA results on beauty production are statistically 
limited, and it is evident that higher luminosity would be beneficial. 
Here we ask more specifically the question whether it will be possible
to vary the two scales $p_T^2$ and $Q^2$ independently and over 
a range extending significantly beyond the one set by the quark mass, 
$4 m_b^2$.
This would open kinematic regions to investigate the regime most relevant 
for improving the perturbative QCD calculation. 
Figure~\ref{fig:heratwo} shows the estimated number of beauty candidates
to be expected if the H1 muon analysis described in section 2 would 
be extended in its present form to the full anticipated HERA II statistics. 
The event yield is displayed as a function of the transverse momentum of 
the $b$ quark in the hadronic centre-of-mass system. 
Reasonable results can be obtained with samples of about 100~events. 
Thus, in photoproduction (DIS) measurements up to 
$p_T\approx 25 (15)$~GeV will be possible. 
In the high $Q^2$ regime, 
more inclusive techniques will be needed to extend the range  
beyond $p_T\approx 10$~GeV.
With the correlation of $x$ and $Q^2$ at HERA in mind, one can 
foresee meaningful tests of low $x$ or high $p_T$ resummation approaches. 

ZEUS is presently being equipped with a
micro-vertex detector~\cite{zeusupgrade}, 
while first results obtained with the H1 silicon tracker 
have been described here.
A major goal is to apply inclusive secondary vertex tagging methods
with algorithms similar to those used at the Tevatron or at LEP, 
where they provided very pure $b$ samples with 
efficiencies of 20~\% and higher. 
Impact parameter resolutions at HERA are similar, and the
luminosity upgrade by means of stronger beam focusing has 
the appreciated side effect of reducing the size of the interaction region
to $80 \times 20 \; \mu\mbox{m}^2$, 
to be compared with  $110 \times 10 \; \mu\mbox{m}^2$, e.g.\  at LEP. 
It is difficult to seriously predict $b$ tagging efficiencies
for HERA, because they depend crucially on how effectively charm and 
light quark background can be suppressed;
excellent control of tracking systematics is a key issue here. 
The method will work the better, the higher $p_T$ and thus the 
vertexing accuracy is. It can complement 
the more exclusive channels in the region 
where they suffer most from limited statistics. 
It will be interesting to check at HERA whether the better description of $b$ 
jet data found in $\pbp$ interactions (figure~\ref{fig:ppxsect}) 
is a general feature. 

Extending the measurements, on the other hand,  
towards low $p_T$ holds the promise of reducing 
the model uncertainties related to the extrapolations into the region where 
the cross section is largest. 
This could be achieved by replacing the semi-leptonic signature 
relying on jets by (semi-)exclusive decay modes involving reconstructed 
$D^{\ast}$ or $J/\Psi$ mesons. 
Due to the higher background levels at lower $p_T$, enhanced discrimination 
power is required online to fully benefit from the higher 
luminosity.  H1 is installing new systems~\cite{h1upgrade}
using latest generation electronics, 
which will allow  to detect invariant mass signatures at the trigger level. 
ZEUS plans to incorporate the  vertex detector into the trigger. 

Finally, both experiments are upgrading their forward tracking systems
with new drift chambers
and silicon detectors~\cite{zeusupgrade,h1upgrade}. 
This will increase the lever arm in $x$ (and $Q^2$) for QCD studies, 
probe the gluon distribution in a different region, up to 
$x\gapprox 0.1$, and give better access to the region where resolved 
photon processes are important. 
Moreover, since at HERA the production of heavy particles,
due to necessarily large $x$ values involved,  
leads to final states boosted into the forward direction, these
upgrades will open possibilities to use beautiful 
signatures in searches for new physics, for example the 
anomalous production of single top quarks, 
for which candidates exist~\cite{heratop}, but not with a $b$ tag.   

In conclusion, in disclosing beauty at HERA, we are entering a field
which is experimentally challenging, theoretically rewarding, 
and potentially exciting. 

\ack
It is a pleasure to thank the organizers for composing 
a stimulating meeting, as unique and irreplaceable as its venue. 
I am indebted to E.~Elsen, L.~Gladilin, B.~Kniehl and M.~Kuze 
for their careful reading of the manuscript.

\end{document}